\documentclass[useAMS,usenatbib]{mn2e}
\usepackage{aas_macros}
\usepackage{amsmath, amssymb}
\usepackage{graphicx}
\usepackage[caption=false,lofdepth,lotdepth]{subfig}
\usepackage{verbatim}

\begin{document}

\title[Measuring the 21-cm difference PDF]{Measuring the History 
of Cosmic Reionization using the 21-cm Difference PDF}
\author[T. Pan, R. Barkana]{Tony Pan$^1$, Rennan Barkana$^2$ \\
$^1$Harvard-Smithsonian Center for Astrophysics, 60 Garden Street,
Cambridge, MA 02138, USA\\ $^2$Raymond and Beverly Sackler School of
Physics and Astronomy, Tel Aviv University, Tel Aviv 69978, Israel}

\pagerange{\pageref{firstpage}--\pageref{lastpage}} \pubyear{2012}

\maketitle

\label{firstpage}
\begin{abstract}

During cosmic reionization, the 21-cm brightness fluctuations were
highly non-Gaussian, and complementary statistics can be extracted
from the distribution of pixel brightness temperatures that are not
derivable from the 21-cm power spectrum.  One such statistic is the
21-cm difference PDF, the probability distribution function of the
difference in the 21-cm brightness temperatures between two points, as
a function of the distance between the points. Guided by 21-cm
difference PDFs extracted from simulations, we perform a maximum
likelihood analysis on mock observational data, and analyze the
ability of present and future low-frequency radio array experiments to
estimate the shape of the 21-cm difference PDF, and measure the
history of cosmic reionization. We find that one-year data with an
experiment such as the Murchison Wide-field Array should suffice for
probing large scales during the mid-to-late stages of reionization,
while a second-generation experiment should yield detailed
measurements over a wide range of scales during most of the
reionization era.

\end{abstract}
\label{lastpage}


\section{Introduction}

In the coming decade, low-frequency radio arrays will begin to probe
the epoch of reionization via the redshifted 21-cm hydrogen
line. Current observational efforts include the MWA (Murchison
Wide-field Array) \citep{MWAref}, LOFAR (Low Frequency Array)
\citep{LOFARref}, PAPER (Precision Array for Probing the Epoch of 
Reionization) \citep{PAPER}, and the GMRT (Giant Metrewave Radio
Telescope) \citep{GMRT}.  Successful interpretation of these
observations will require effective statistical techniques for
analyzing the data. Due to the difficulty of these measurements, it is
important to develop techniques beyond the standard power spectrum
analysis, in order to offer independent confirmation of the
reionization history, probe different aspects of the topology of
reionization, and do this with methods subject to different systematic
errors.

During reionization, the hydrogen distribution is a highly non-linear
function of the distribution of the underlying ionizing sources. A
natural statistic for probing the expected non-Gaussianity is the
one-point probability distribution function (PDF) of the 21-cm
brightness temperature at a point
\citep{fzh04,ciardi,mellema,wyithe,Harker,Ichikawa2010,Gluscevic2010}.
In this paper, we focus on the PDF $p_\Delta(\Delta T_b)$ of the
difference in 21-cm brightness temperature between two points in the
cosmological volume, $\Delta T_b = |T_2-T_1|$.  This 21-cm difference
PDF, suggested by
\citet{Barkana2008}, is a two-dimensional function, dependent not only
on $\Delta T_b$ but also on the separation $r$ between the two points
at which the difference in brightness temperature is measured.

There are at least three advantages to the 21-cm difference PDF
statistics \citep{Gluscevic2010}, which we summarize here.  Firstly,
if the number of resolved cubic pixels (i.e., voxels) in the observed
volume is $N$, the number of data points available for reconstructing
the one-point PDF is only $N$, whereas the number of data points
available for reconstructing the difference PDF ($N^2/2$) is
overwhelmingly larger, albeit the latter data points must be sorted
into bins of distance $r$. Thus, we might expect to do better than
with the one-point PDF, which requires rather strong assumptions in
order to allow a reconstruction of the reionization history with
upcoming experiments \citep{Ichikawa2010}. Secondly, the 21-cm
difference PDF generalizes both the one-point PDF and the two-point
correlation function of $T_b$ (the latter of which can be deduced
using the variance of the difference PDF, and is equivalent to the
power spectrum), and also yields additional information beyond those
statistics.  Thirdly, the difference PDF avoids (by its very
definition) the unwanted contribution of the mean sky background
temperature, and is readily applicable to temperature differences measured
with radio interferometry.

\section{Methodology}

We adopt the expected parameters for 1-year observations of a single
field of view with MWA, using equations for 21-cm interferometer
arrays from the review by \citet{Furlanetto2006}, with an integration
time of $t_{int}= 1000$ hours, a collecting area of $A_{tot} \sim 2
\times 10^3$ m$^2$, a field of view of $\pi 16^2$ deg$^2$ and a total
bandwidth of $\Delta \nu_{\rm tot} = 6$ Mhz.  Note that the collecting
area here is 4 times smaller than the collecting area assumed in
\citet{Ichikawa2010}, given the scaling down of the first
generation of the MWA compared to earlier plans. Then, assuming cubic
pixels of size $r_{\rm com}$ (all distances comoving), we find the
following expected number of voxels $N_p$ and root-mean-square noise
in each one $\sigma_N$:
\begin{eqnarray}
N_p &=& 8.2\times 10^7 \left( \frac{r_{\rm com}}{2.9\, \mbox{Mpc}}
\right)^{-3} \left( \frac{1+z}{9} \right)^{0.9}\ , \label{Eq_Np}\\
\sigma_N &=& 4400 \left( \frac{r_{\rm com}}{2.9\, \mbox{Mpc}} 
\right)^{-2.5} \left( \frac{1+z}{9} \right)^{5.25} \mbox{mK}\ .
\label{Eq_sigmaN}
\end{eqnarray}

In order to look a bit ahead, we also consider specifications with
lower noise in the same field of view, e.g., 1/2 the noise we denote
as MWA/2 (which corresponds to 4-year data with the MWA), while 1/10
the noise we denote as MWA/10; the latter is a conservative estimate
(by at least a factor of a few) for larger, second generation 21-cm
arrays such as the SKA (Square Kilometer Array) or a 5000-antenna
MWA. The only source of noise we consider is Gaussian thermal noise,
whose magnitude is determined by the receiver's system temperature,
which is set by the sky's brightness temperature dominated by Galactic
synchrotron emission
\citep{Furlanetto2006}. This assumes perfect foreground removal from
21-cm maps. Clearly, the first step for any proposed measurement
method is to prove its feasibility against thermal noise, which can
then motivate more detailed analyses that include a larger range of
observational difficulties and sources of noise.

\subsection{Model PDF and Thermal Noise}

We begin by considering a general PDF, which could be the regular
(one-point) PDF or the difference PDF.  To determine a best-fit PDF
using observed data, we characterize the PDF with a finite number of
parameters.  We do so with a binned PDF, expressed as the sum of
boxcar functions for each bin $p(x)=\sum F_i(x)$, where
\begin{equation}
F_i(x) = \left\{ 
  \begin{array}{l l}
    l_i & \quad \textrm{if $a_i<x<a_{i+1}$}\\
    0 & \quad \textrm{otherwise.}\\
  \end{array} \right.
\end{equation}
Here $l_i$ is the bin height, and the probability contained in bin $i$
is $p_i = l_i(a_{i+1}-a_i)$.  In a model with $N_b$ bins where the bin
edges $\{a_1,a_2,\ldots,a_{N_b+1}\}$ are fixed, the binned PDF only
has $N_b-1$ free parameters $\{l_1,l_2,\ldots,l_{N_b-1}\}$, as
$l_{N_b}$ must be normalized such that the probabilities sum up to
unity. Note that throughout this paper we use this very general binned
form for the PDF, and do not need to assume a particular functional
form for the difference PDF as is necessary for the one-point PDF
given its much lower signal-to-noise ratio \citep{Ichikawa2010}.

Theoretically, the measured PDF will be the true PDF convolved with
the noise. For example, the 21-cm brightness temperature one-point PDF
measured by instruments will be the true one-point PDF $p(T)$
convolved with an extremely broad normal distribution:
\begin{equation}
N(0,\sigma_N^2) =
\frac{1}{\sqrt{2\pi}\,\sigma_N}\, e^{-\frac{T^2}{2\sigma^2_N}}\ ,
\end{equation} 
with zero mean and standard deviation $\sigma_N$ due to thermal noise
(Equation~(\ref{Eq_sigmaN})). Hence, the noisy one-point PDF can be
expressed as a sum of the convolution of boxcar functions with the
Gaussian,
\begin{equation}
p_{\rm noisy}(T)=\sum F_i(T) \star N(0,\sigma_N^2)\ , 
\end{equation}
where
\begin{equation}
F_i(T) \star N(0,\sigma_N^2) = \frac{1}{2} l_i \left[
\mbox{Erf}\left(\frac{T-a_i}{\sqrt{2}\,\sigma_N}\right) -
\mbox{Erf}\left(\frac{T-a_{i+1}}{\sqrt{2}\,\sigma_N}\right) \right].
\end{equation}
Here $\mbox{Erf}(x)$ is the error function.  $p_{\rm noisy}(T)$ can be
used to generate mock observations of the one-point PDF.

Out of symmetry and convenience, the difference PDF $p_\Delta(\Delta
T_b)$ defined in the literature \citep{Barkana2008} is a function of
the \emph{absolute} difference in brightness temperature $\Delta T_b =
|T_2-T_1|$.  However, to find the effect of thermal noise on the
difference PDF $p_{\Delta,\rm noisy}(\Delta T_b)$, it is easier to first find
the probability distribution function $p_{\Delta,\rm noisy}(\Delta T_{21})$
as a function of the temperature difference $\Delta T_{21}
\equiv T_2-T_1$ without the absolute value.  If the thermal noise at
any two points 1 and 2 is uncorrelated, then their temperature
difference has a root-mean-square thermal noise of $\sqrt{2}\,\sigma_N$.
If the intrinsic difference PDF is $p_\Delta(\Delta T_{21})$, the
observed version is then:
\begin{eqnarray}
\nonumber
&& p_{\Delta,\rm noisy}(\Delta T_{21}) \\
\nonumber
&=& p_\Delta(\Delta T_{21}) \star N(0,2\sigma_N^2) \\
\nonumber
	&=& \sum F_i(\Delta T_{21}) \star N(0,2 \sigma_N^2) \\
	&=& \sum \frac{1}{2} l_i \left[  \mbox{Erf}\left(\frac{\Delta 
T_{21}-a_i}{2\sigma_N}\right) - 
\mbox{Erf}\left(\frac{\Delta T_{21}-a_{i+1}}{2\sigma_N}\right) \right].
\label{Eq_p_noise_T21}
\end{eqnarray}
Now, due to symmetry, $p_{\Delta,\rm noisy}(\Delta T_{21})$ is an even
function, so we can recover the difference PDF as defined for the
absolute temperature differences, with thermal noise included, via
\begin{equation}
p_{\Delta,\rm noisy}(\Delta T_b) = \left\{ 
  \begin{array}{l l}
    2\: p_{\Delta,\rm noisy}(\Delta T_{21}) & 
\quad \textrm{if $\Delta T_{21}\ge0$}\\
    0 & \quad \textrm{otherwise}\ .\\
  \end{array} \right. 
\end{equation}

To use equation~(\ref{Eq_p_noise_T21}), we need to assume a true
difference PDF, with which to convolve the thermal noise.  To this
end, we use the binned difference PDF as measured in the fiducial $S1$
simulation of \citet{McQuinn2007}, who modeled the density field
during the epoch of reionization with a $1024^3$ N-body simulation in
a box size of $\approx 94$ Mpc, post-processing it using a suite of
radiative-transfer simulations to characterize the morphology and size
distribution of ionized regions during reionization. Analytic
prescriptions were used to model reionization effects of small-scale
structure that was unresolved in the N-body simulation. Source
parameters were chosen so that reionization ends near $z=7$ in the
simulation.

\begin{figure*}
\centering
\subfloat{\includegraphics[width=0.31\textwidth]{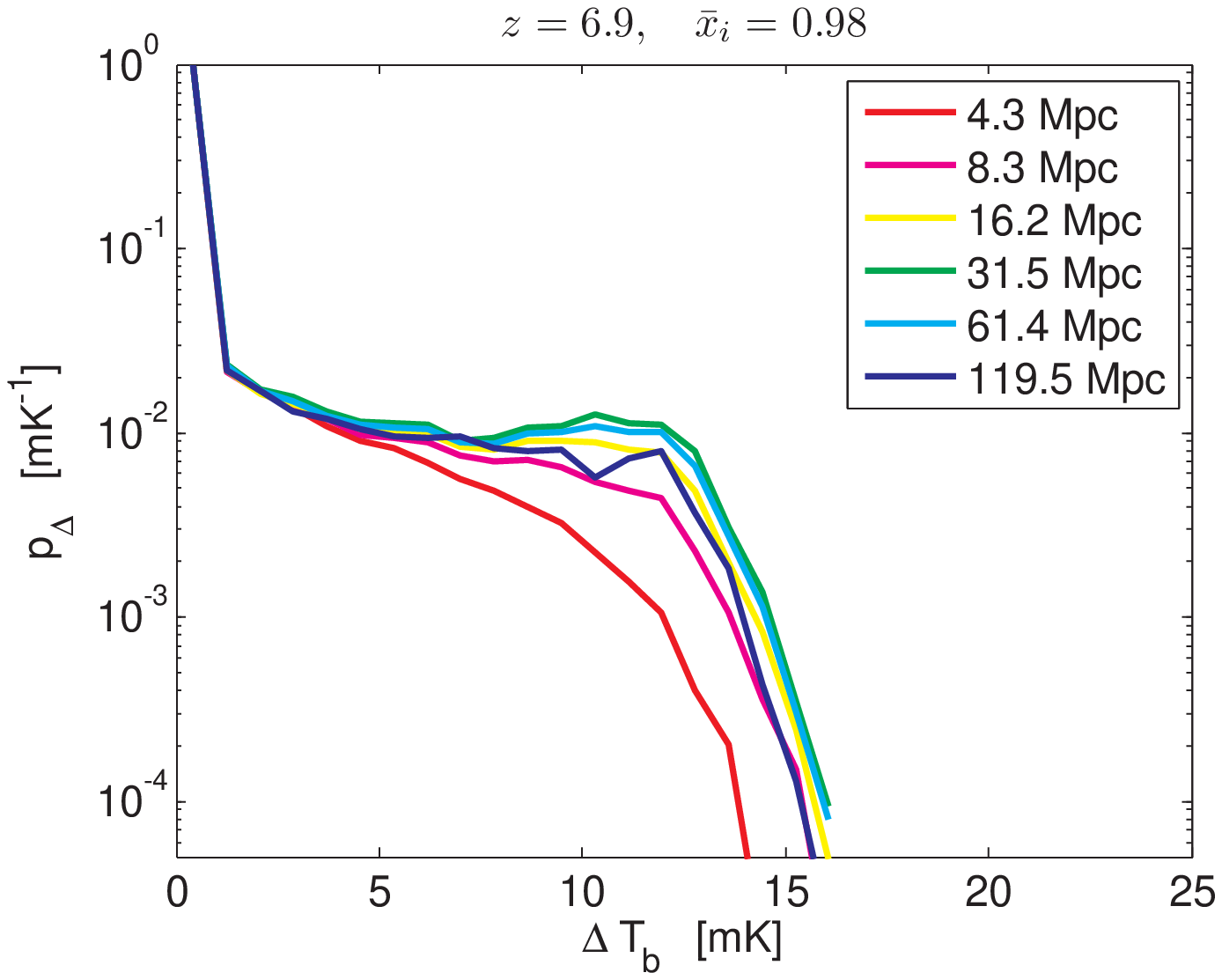}}
\quad
\subfloat{\includegraphics[width=0.31\textwidth]{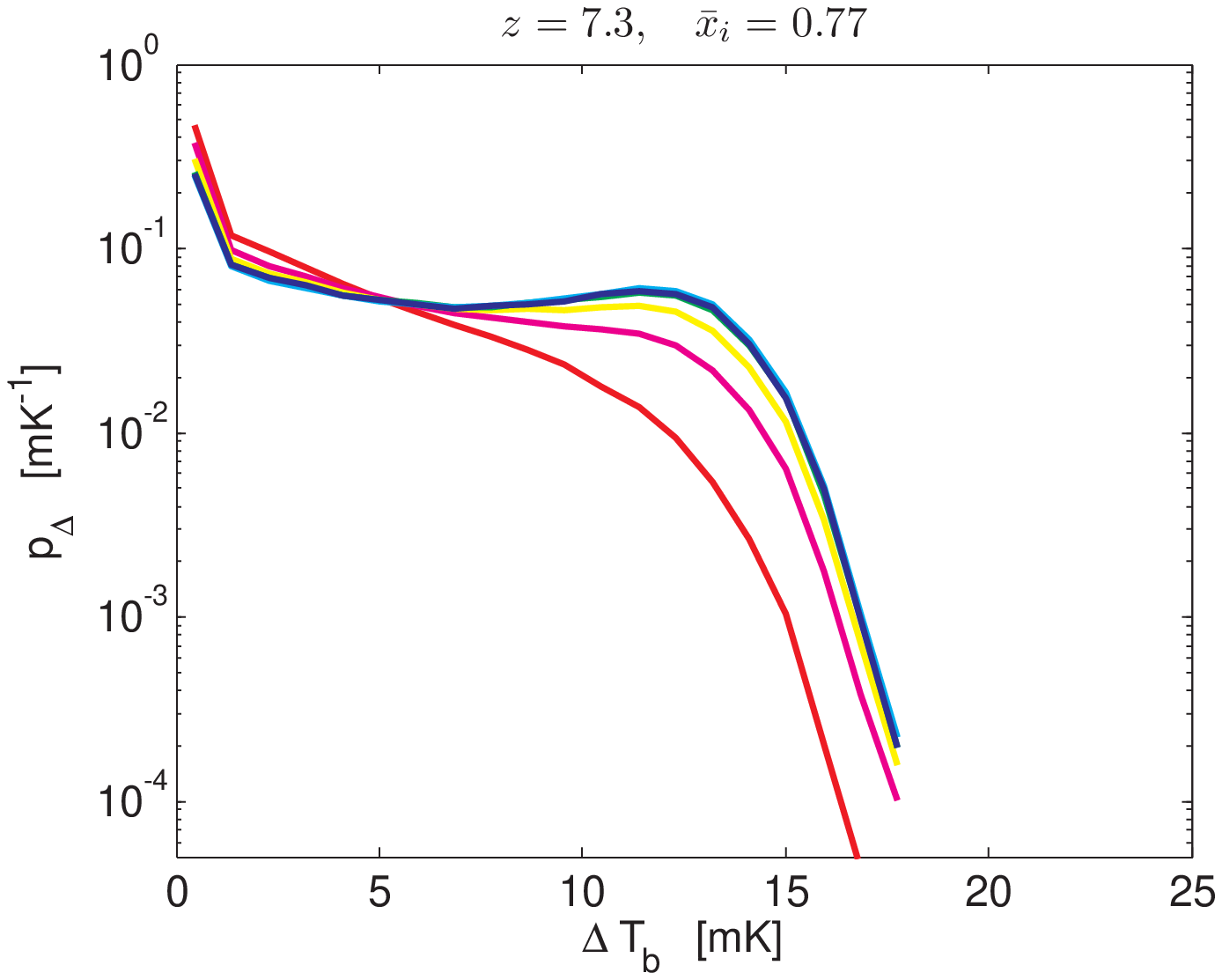}}
\quad
\subfloat{\includegraphics[width=0.31\textwidth]{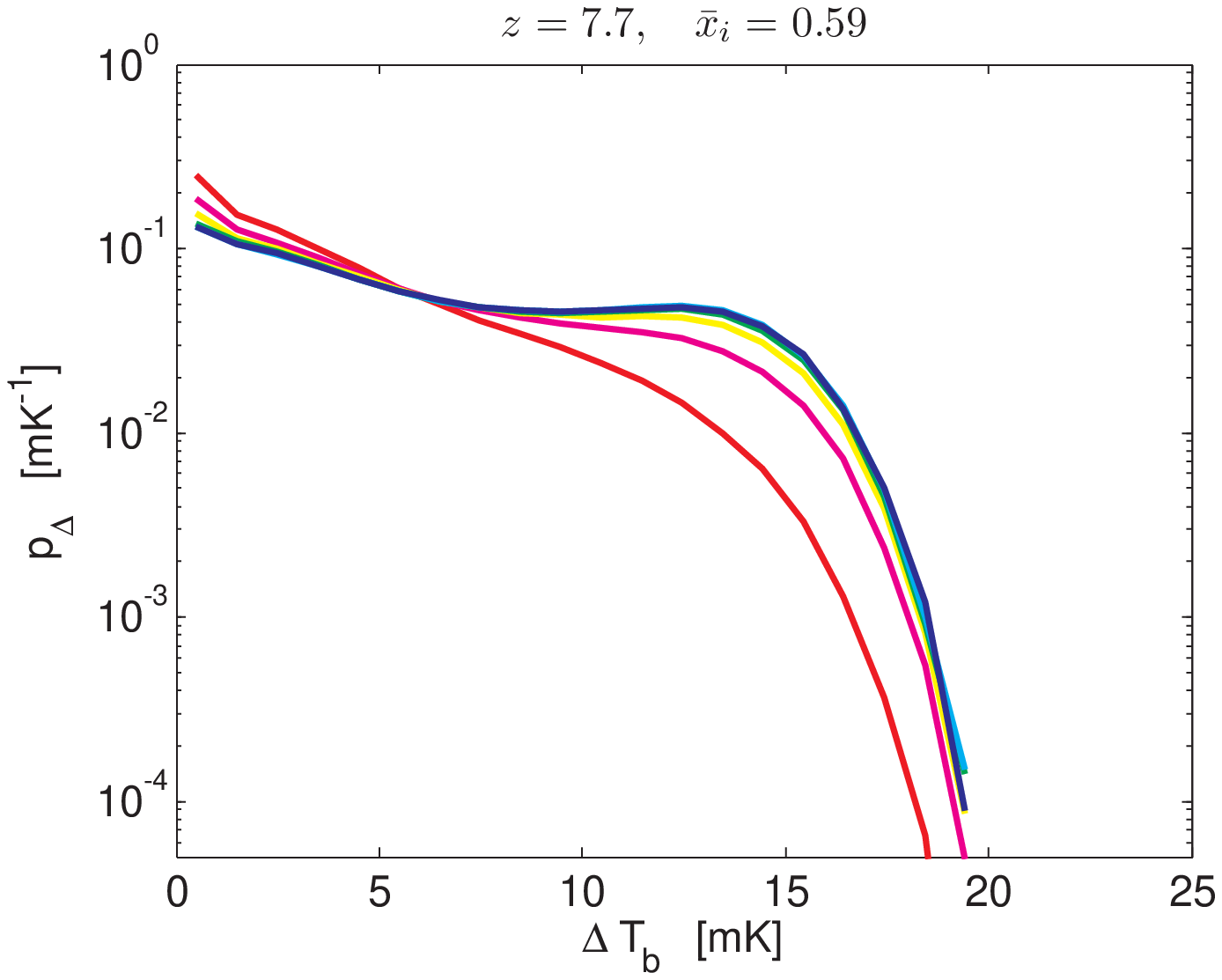}}
\\
\subfloat{\includegraphics[width=0.31\textwidth]{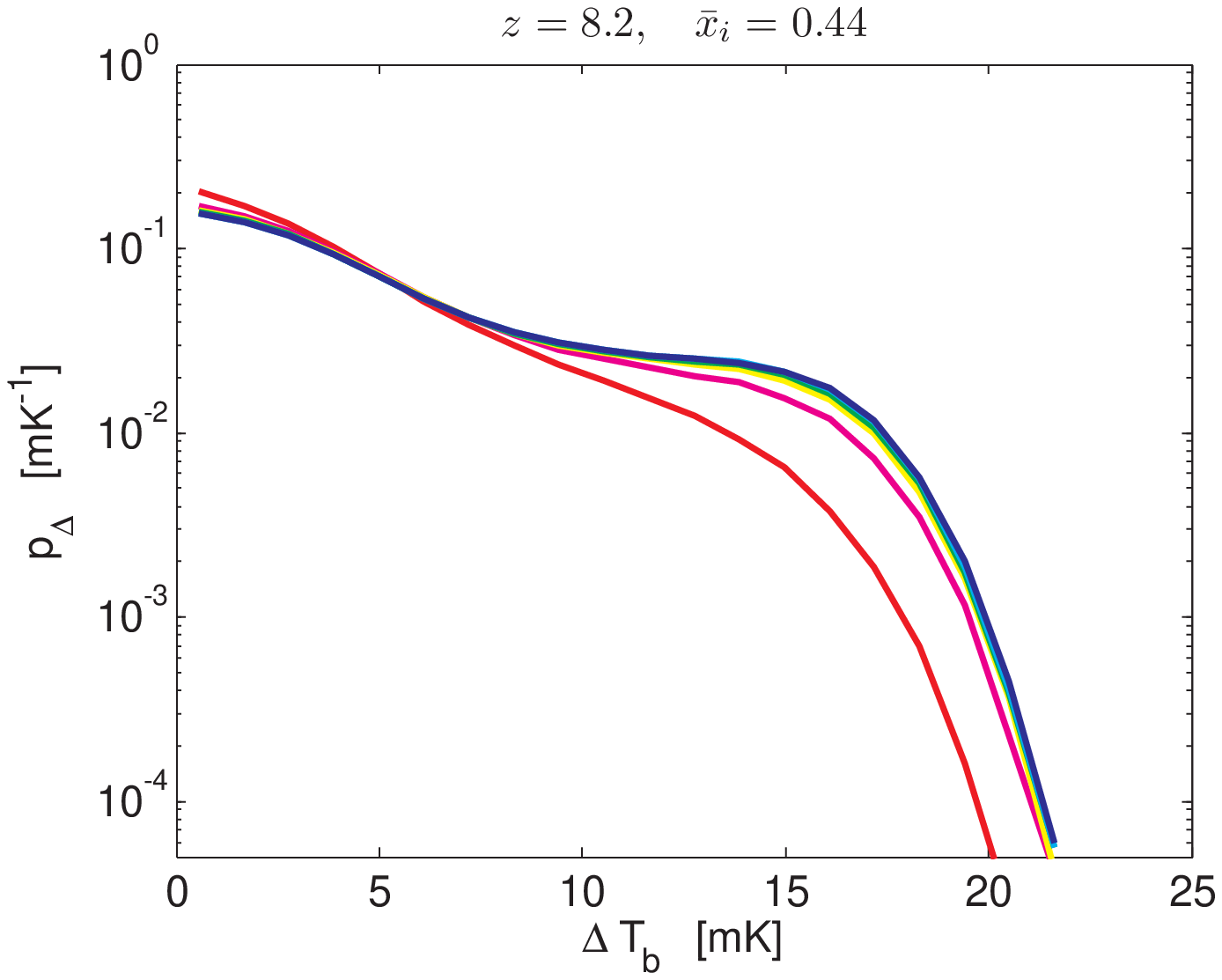}}
\quad
\subfloat{\includegraphics[width=0.31\textwidth]{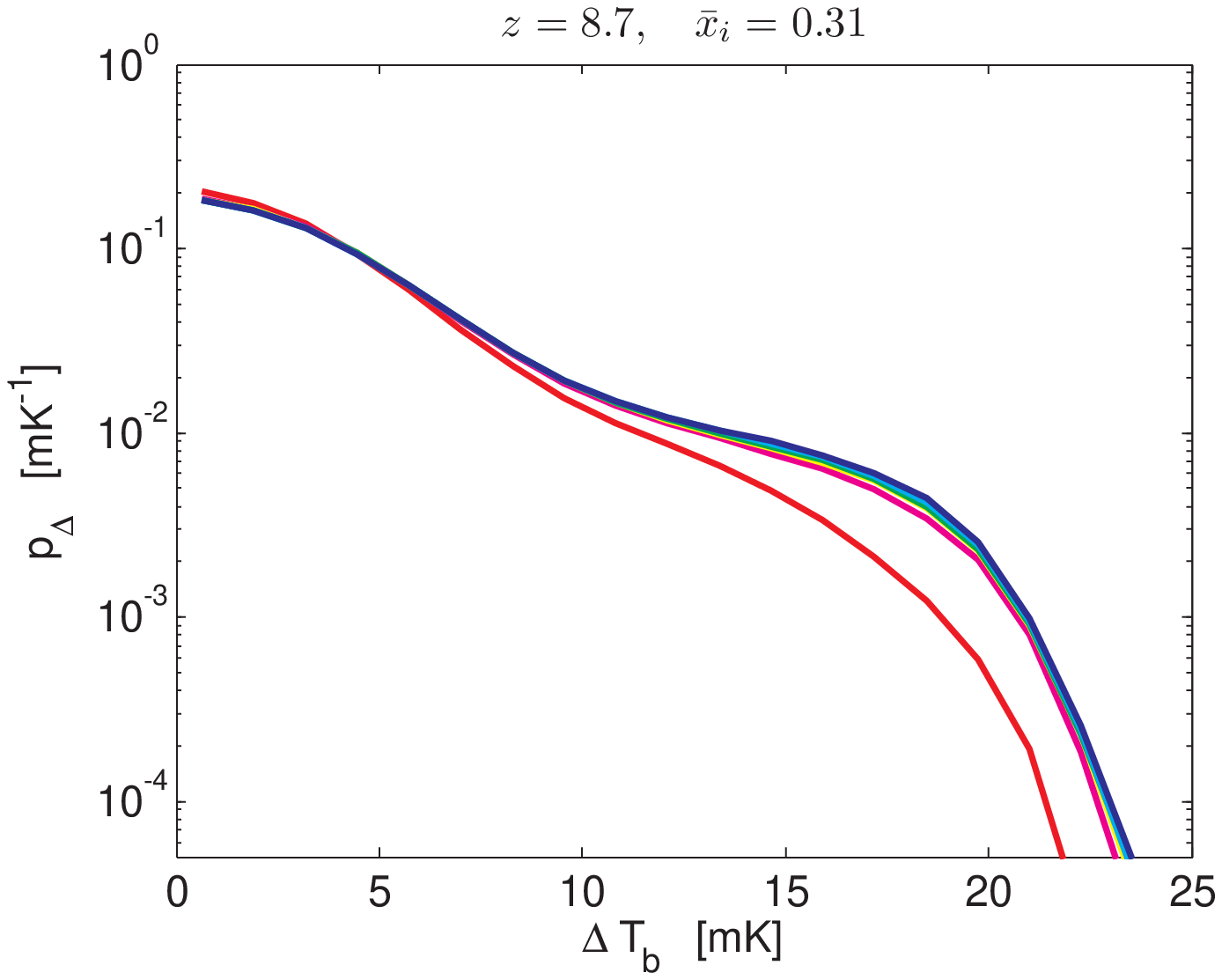}}
\quad
\subfloat{\includegraphics[width=0.31\textwidth]{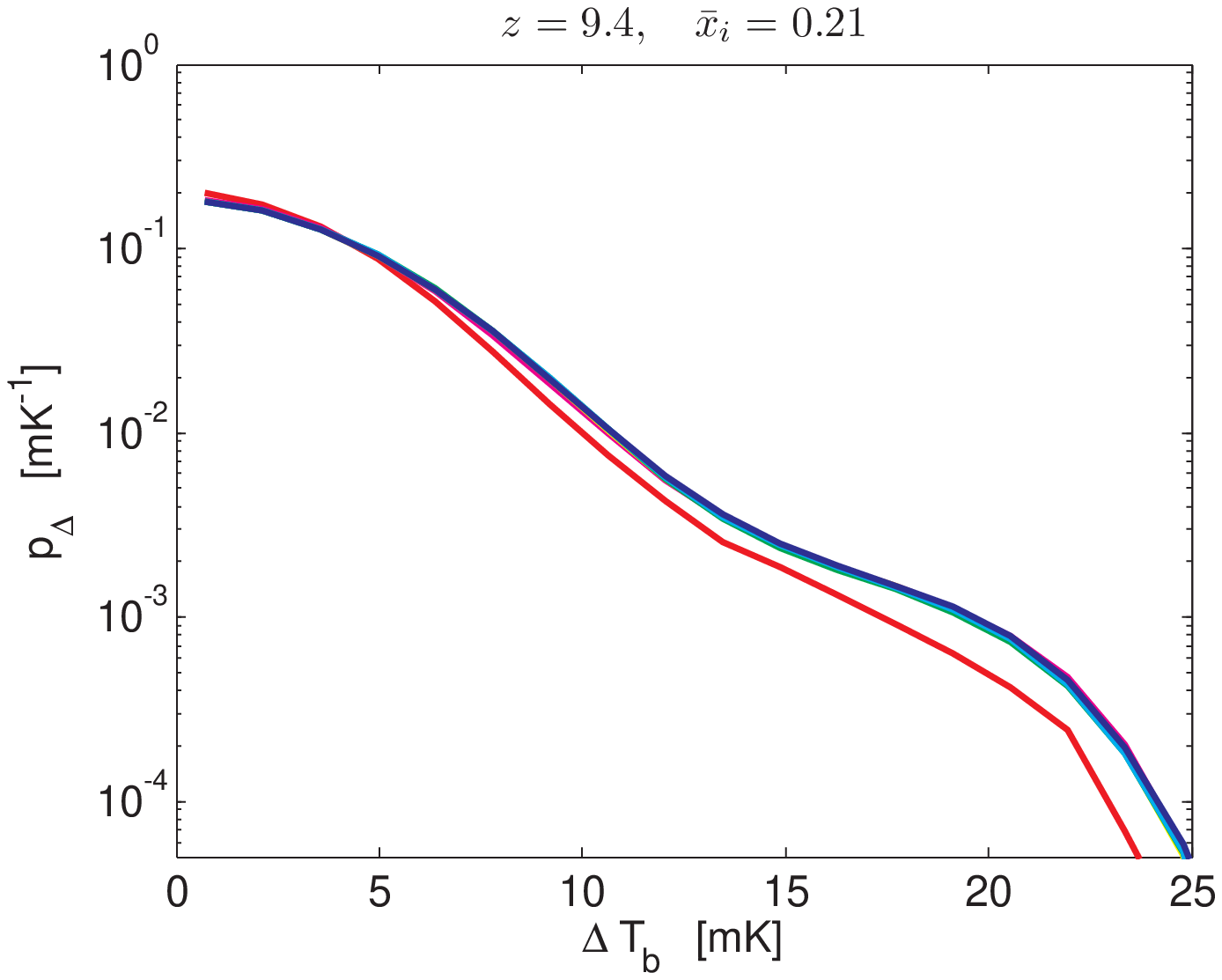}}
\\
\subfloat{\includegraphics[width=0.31\textwidth]{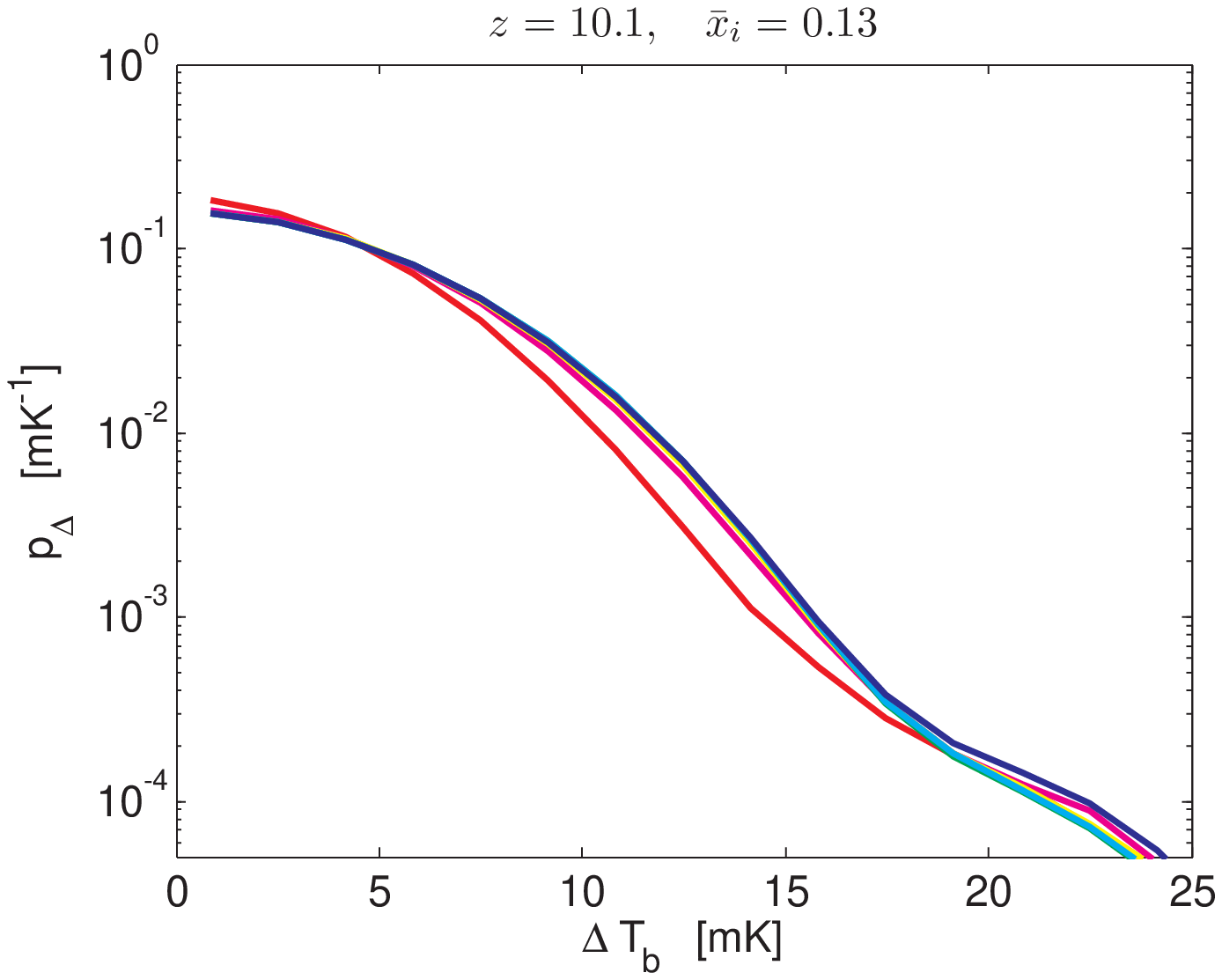}}
\quad
\subfloat{\includegraphics[width=0.31\textwidth]{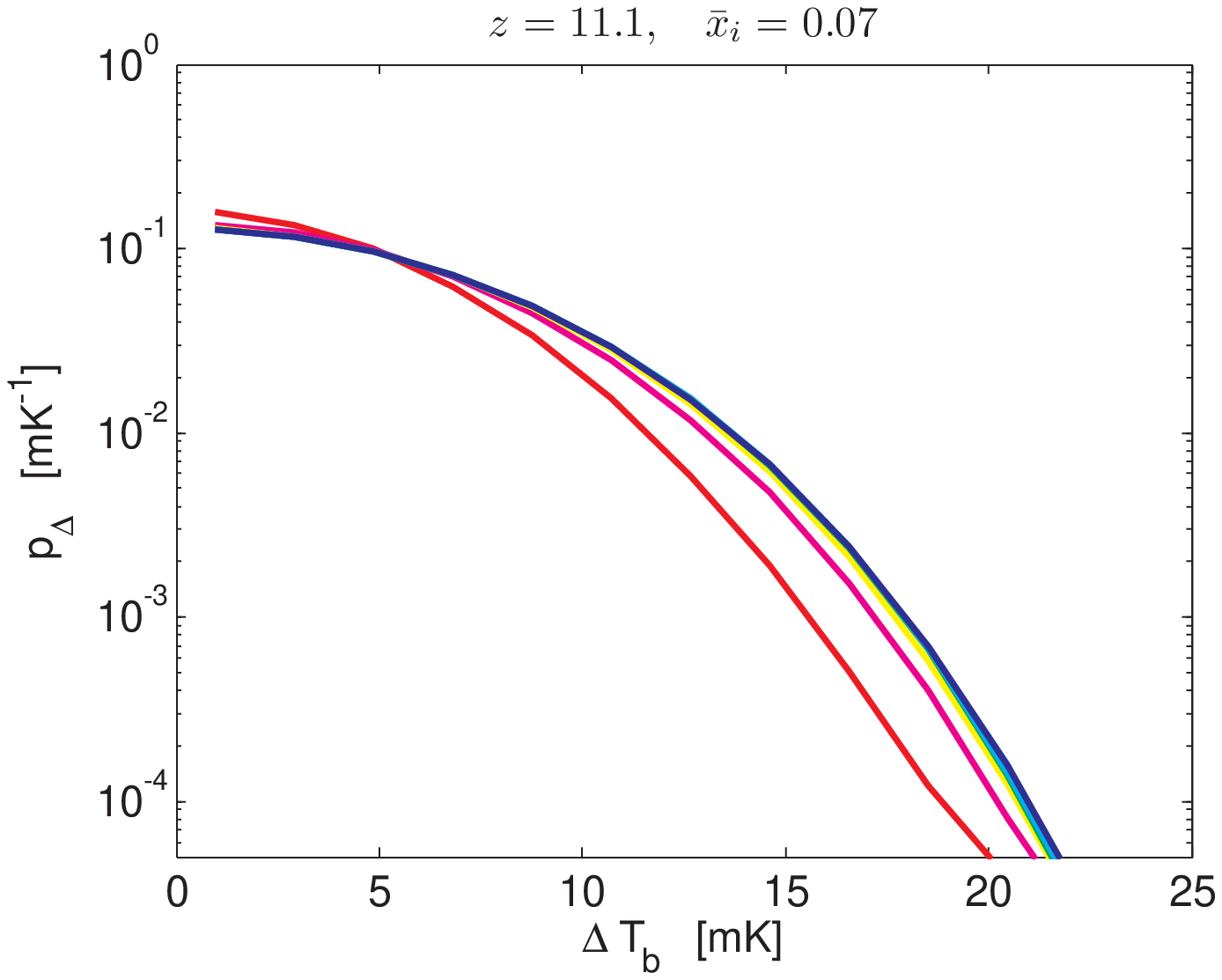}}
\quad
\subfloat{\includegraphics[width=0.31\textwidth]{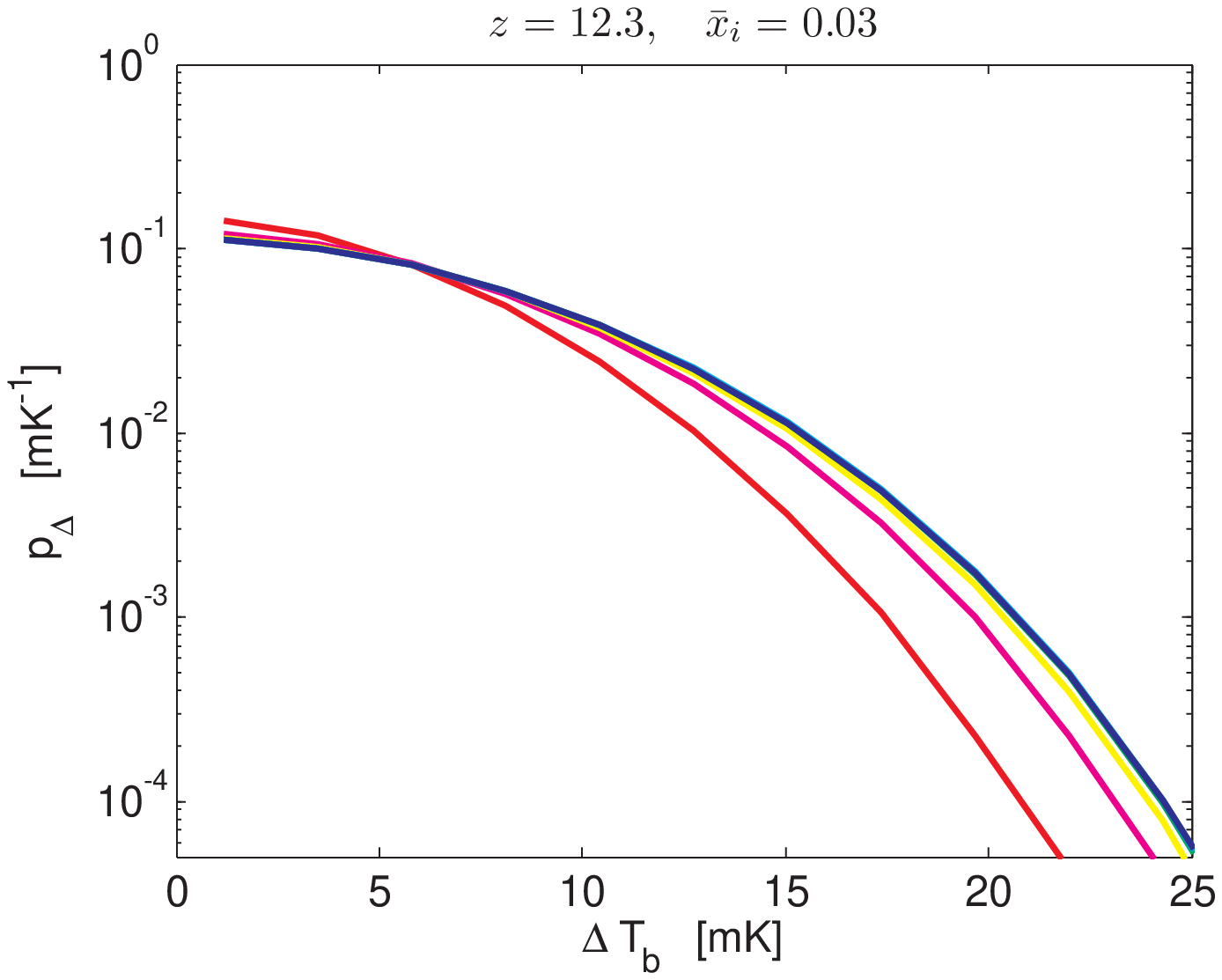}}
\caption{21-cm difference PDFs from \citet{Gluscevic2010} are shown 
here as a function of their distance bin and redshift.  The legend in
the first panel indicates the central values $r_{\rm mid}$ of the
logarithmic distance bins.}
\label{TrueSimulatedPDF}
\end{figure*}

The 21-cm difference PDF's from these simulations, shown in
Figure~\ref{TrueSimulatedPDF}, were first presented by
\citet{Gluscevic2010}; here we use the same redshift slices (taken at
50 Myr intervals), the same cubic voxel size of 2.9 (comoving) Mpc,
and the same logarithmically spaced distance bins to obtain the `true'
difference PDF from the simulation.  The central values of the
logarithmic distance bins are $r_{\rm mid}=$ 4.3, 8.3, 16.2, 31.5,
61.4, and 119.5 Mpc.  However, instead of using the same 20 linearly
spaced temperature bins as \citet{Gluscevic2010}, we alternate the
number and interval size of our temperature bins
$\{a_1,a_2,\ldots,a_{N_b+1}\}$, to see the dependence of the fit
errors on the number of free parameters. In general, reducing the
number of bins gives a more accurate determination of the PDF, but at
the price of less detailed information on its shape, since the
measured PDF is (at best) the true one but smoothed on the scale of
the bin size.

Mock observational difference PDF values can be created by randomly
generating $n$ values of $\Delta T_b$ using the distribution
$p_{\Delta,\rm noisy}(\Delta T_b)$.  The same functional form can be employed in
finding the best fit parameters $\{l_i\}$ to the same mock
observations with a maximum likelihood method.  In this paper, we
sample each difference PDF and generate 1000 Monte Carlo instances of
observational data for that model, and thus obtain a well-sampled
distribution of reconstructed model parameters.

\subsection{Number of Voxel Pairs}

For a given observational volume, voxel size, and distance $r$ between
pairs of voxels, the number of voxel-pairs $N(r) dr$ (at distances
between $r$ and $r+dr$) is uniquely determined.  Thus, for each bin we
sample $n=\int N(r) dr$ values of $\Delta T_b$ from $p_{\Delta,\rm
noisy}(\Delta T_b)$ as our mock data.

With the edge size of each voxel normalized to 1, the number of
voxel-pairs in a cubic volume $V=L^3$ as a function of the voxel
distance $r$ can be closely approximately by the number of voxel-pairs
in a sphere of the same volume $V=4\pi R^3/3$ at the same distance, at
least for small $r$. A sphere is easier to analyze, and yields an
analytical result for the voxel-pair distribution function:
\begin{equation}
N(r)\, dr = \frac{1}{6}\pi^2 r^2 (2R-r)^2 (4R+r)\, dr\ .
\label{Eq_Nr}
\end{equation}
Equation~(\ref{Eq_Nr}) is exact for spherical volumes for all
$r\in(0,2R)$, in the limit of infinitesimal $dr$ and voxel size
(compared to $r$ and $R$). The total number of voxel pairs at all $r$
in a sphere of radius $R$ is
\begin{equation}
\int_{0}^{2R} N(r)\, dr = \frac{1}{2} \left( \frac{4}{3}\pi R^3 \right)^2,
\end{equation}
which is $\frac{1}{2}N^2$ as expected, in terms of the total number of
voxels $N$. 

If the distance between voxel-pairs is much less than the
characteristic size of the observable volume, then the total number of
voxel-pairs at that small distance will not be sensitive to the shape
of the volume, but only to its size. In considering such a pair, voxel
\#1 can be chosen anywhere within the volume, and \#2 must
then be at a distance $r$ in any direction. As long as $r \ll L$ (or
$r \ll R$), the full sphere of radius $r$ about \# 1 will almost
always fall within the big volume, so that all \#2 voxels on this
small sphere are allowed. Thus, the number of pairs will be the number of 
\#1 voxels $V$ times the number of \#2 voxels $4 \pi r^2\, dr$, 
divided by 2 for double-counting of pairs. The result of $2 \pi V
r^2\, dr$ agrees with equation~(\ref{Eq_Nr}) in the small-$r$
limit. Corrections to this result will come from cases where pixel \#1
is within a distance $r$ of the volume's boundary, i.e., the
correction is of order the surface area times $r$ divided by the
volume, which (for simply-connected convex volumes such as a sphere or
cube) is of order $r/V^{1/3}$. 

For the MWA, if we assume that the volume it will observe on the sky
is approximately cubic, equation~(\ref{Eq_Np}) implies that the length
of the cube $L \sim 1000$ Mpc for all redshifts of interest.  This is
much greater than the largest distance ($r \sim 163$ Mpc) between
voxel-pairs we consider in this paper.  Since $L\gg r$, we use
equation~(\ref{Eq_Nr}) as an excellent proxy for the number of
voxel-pairs MWA will observe, and list the values of $\int N(r) dr$ in
each distance bin in Table~\ref{VoxelPairTable}. 

\begin{table}
 	 \caption{ The number of voxel-pairs $\int N(r) dr$ in each
 	 distance bin, as a function of redshift.  The distance bins
 	 are logarithmically spaced and are denoted via their central
 	 values $r_{\rm mid}$, in units of (comoving) Mpc.  The size
 	 of each voxel is chosen to be 2.9 Mpc, consistent with
 	 \citet{Gluscevic2010}, from which the simulated difference
 	 PDFs were taken.  Due to equation~\ref{Eq_Np}, the total
 	 number of voxels (and thus $N(r)$) has a slight redshift
 	 dependence.  Higher separation bins $r_{\rm mid}$ have orders
 	 of magnitude more voxel-pairs compared to lower $r_{\rm
 	 mid}$, varying roughly as $r^3$ when $r\ll L$ (since the bin
 	 width $\propto r$).}  \begin{tabular}{ | l | l | l | l | l |
 	 l | l |} \hline $r_{\rm mid}=$ & 4.3 & 8.3 & 16.2 & 31.5 &
 	 61.4 & 119.5 \\ \hline $z=6.9$ & 1.1E9 & 8.3E9 & 6.1E10 &
 	 4.4E11 & 3.1E12 & 2.2E13 \\ $z=7.3$ & 1.2E9 & 8.7E9 & 6.4E10
 	 & 4.6E11 & 3.3E12 & 2.3E13 \\ $z=7.7$ & 1.2E9 & 9.1E9 &
 	 6.6E10 & 4.8E11 & 3.4E12 & 2.4E13 \\ $z=8.2$ & 1.3E9 & 9.6E9
 	 & 7.0E10 & 5.1E11 & 3.6E12 & 2.5E13 \\ $z=8.7$ & 1.4E9 &
 	 1.0E10 & 7.3E10 & 5.3E11 & 3.8E12 & 2.6E13 \\ $z=9.4$ & 1.5E9
 	 & 1.1E10 & 7.8E10 & 5.7E11 & 4.1E12 & 2.8E13 \\ \hline
 	 \end{tabular}
\label{VoxelPairTable}
\end{table}

\subsection{Maximum Likelihood for a Multinomial Distribution}

Since $n$ is large, to compare the mock observational data to a
potential model, we also bin the observational data into $N_B$ bins
(in general different from the number of bins $N_b$ in the model).
Note that a binned PDF is essentially a multinomial distribution of
the variable $X$ given by the set of bin probabilities
$\mathbf{p}=(p_1,\ldots ,p_{N_B})$; given $n$ total data points, there
will be an expected number of $n_{\mbox{exp},j}=n\: p_j$ data points
in bin $j$.  As for the covariance matrix $\Sigma$, the variance of
$X$ in a single bin $j$ is $p_j(1-p_j)$, while the covariance between
different bins $i$, $j$ is $-p_i p_j$.  How does one account for the
covariance structure of a multinomial distribution in a maximum
likelihood estimate (MLE) fit?

In the limit of large $n$, the multinomial distribution is
approximated by the multivariate normal distribution with the same
mean $\mathbf{p}$ and covariance $\Sigma$. We apply MLE to this
multivariate normal with model parameters
$\mathbf{p}^{\ast}=(p_1,\ldots,p_{N_B-1})$, where we drop the last bin
$p_{N_B}$ because it is not an independent variable due to
normalization constraints, and none of the elements in $\Sigma$ are
free variables as they are completely determined by
$\mathbf{p}^{\ast}$. We thus find that the objective function to be
minimized is:
\begin{equation}
(\Delta X^{\ast})^T (\Sigma^{\ast})^{-1} (\Delta X^{\ast}),
\end{equation}
where $X^{\ast}$ refers to the first $N_B-1$ bins, so that
the vector
\begin{eqnarray}
\nonumber
\Delta X^{\ast} 	&=& \overline{X^{\ast}}-\mathbf{p}^{\ast} \\
	&=& (\overline{X_1}-p_1,\ldots,\overline{X_{N_{B-1}}}-p_{N_B-1})
\end{eqnarray}
is the deviation between the observed probabilities
$\overline{X^{\ast}}$ and model probabilities $\mathbf{p}^{\ast}$.
Note that by definition $\overline{X_j}=n_j/n$, where $n_j$ is the
actual number of data points observed in bin $j$.  Similarly,
$\Sigma^{\ast}$ is the covariance matrix of $X^{\ast}$, and is equal
to the upper-left $(N_B-1)\times (N_B-1)$ submatrix of $\Sigma$.
Note that this approach correctly accounts for the constraint of
the total probability summing up to unity.

$\Sigma^{\ast}$ is indeed invertible, and takes the form:
\begin{equation}
 (\Sigma^{\ast})^{-1} = \left( 
\begin{array}{cccc}
\frac{1}{p_1}+\frac{1}{p_{N_B}}	& \frac{1}{p_{N_B}}		
				& \ldots		
	& \frac{1}{p_{N_B}}	  \\
\frac{1}{p_{N_B}}					
	& \frac{1}{p_2}+\frac{1}{p_{N_B}}	& \ldots		
	& \frac{1}{p_{N_B}}	  \\
\vdots								
	& \vdots								
		& \ddots 		
& \vdots					  \\
\frac{1}{p_{N_B}}					
	& \frac{1}{p_{N_B}}					
	& \ldots			
& \frac{1}{p_{N_B-1}}+\frac{1}{p_{N_B}}	  
\end{array} 
\right)\ . \end{equation}
Thus, we find that the objective function
\begin{eqnarray}
\nonumber
&&		(\Delta X^{\ast})^T (\Sigma^{\ast})^{-1} (\Delta X^{\ast}) \\
\nonumber
&=& \left(
\frac{\overline{X_1}}{p_1}-\frac{X_{N_B}}{p_{N_B}},
\ldots,\frac{\overline{X_{N_{B-1}}}}{p_{N_{B-1}}}-\frac{X_{N_B}}{p_{N_B}}
\right)^T (\Delta X^{\ast}) \\
\nonumber
&=& \sum_{j=1}^{N_B} \frac{(\overline{X_j}-p_j)^2}{p_j} \\
\nonumber
&=& \frac{1}{n} \sum_{j=1}^{N_B}
\frac{(n_j-n_{\mbox{exp},j})^2}{n_{\mbox{exp},j}} \\ &=& \frac{1}{n}
\chi^2,
\end{eqnarray}
where $\chi^2$ is the standard Pearson's chi-squared statistic.  In
summary, to find the best MLE fit for a multinomial distribution of
$N_B$ components with $N_B-1$ free parameters, one can simply minimize
a standard $\chi^2$ statistic in which \emph{all} $N_B$ terms in the
$\chi^2$ are summed over.

We bin the values from the mock observation data into $N_B=1,000$
bins; this is justified as long as the bin width is much smaller than
any scale we hope to resolve in the observed PDF.  We leave the last
bin at the tail of the observed difference PDF with a much wider width
than the other bins, so that each bin has more than 10 counts (with
most having a vastly larger count number), and we do not have to use
the $C$-statistic \citep{Cash1979} instead to account for large
relative errors at small counts.

\section{Results}

\subsection{1-bin model}

The attempt to measure the difference PDF is essentially a contest
between a very high level of noise per measurement (almost three
orders of magnitude larger than the width of the intrinsic difference
PDF in Figure~\ref{TrueSimulatedPDF}) and a very large number of
measurements. A naive signal-to-noise estimate may suggest that only
$\sim 10^6$ measurements (i.e., the square of the noise-to-signal
ratio of each measurement) are needed for a rough measurement of a
given difference PDF value (in some bin of temperature difference).
In reality, though, the needed number is significantly higher, because
of the near-degeneracy that is encountered in what is essentially an
attempt to deconvolve the noisy difference PDF (see
\citet{Ichikawa2010} for a detailed discussion in the context 
of the one-point PDF). Thus, it is prudent to start out
conservatively, and try to fit a small number of bins.

Luckily, theory suggests that even a 1-bin model is worth considering,
since it can yield valuable information. Such a model consists of a
single bin at $\Delta T_b\approx 0$, plus a normalization bin at
higher values.  of $\Delta T_b$. The value of the total probability
$p_1$ of the difference PDF in this first bin can be used as an
approximation of $\Delta P_D$, which refers to the theoretical limit
of a component of $\Delta P$ which is a Dirac Delta function at
$\Delta T_b = 0$
\citep{Gluscevic2010}. This $\Delta P_D$ effectively measures a low-resolution
version of the ionization correlation function during cosmic
reionization \citep{Barkana2008}; in the limit of perfect resolution,
$\Delta P_D$ would exactly correspond to the joint ionization
probability of two points as a function of their distance $r$. In this
limit, at $r\rightarrow 0$, $\Delta P_D$ should simply equal the
probability of having $T_b=0$ mK, which is the mean ionized fraction
$\overline{x}_i$, and $\Delta P_D$ should decrease with increasing
$r$ as the pair-correlation drops, until $\Delta P_D
\approx \overline{x}_i^2$ at $r\rightarrow\infty$ (for which each voxel in the
pair is ionized independently). As long as the ionized regions
maintain a low (even if non-zero) neutral fraction, this description should
be approximately valid even for the realistic difference PDF. 

\begin{figure}
\centering
\includegraphics[width=1\columnwidth]
{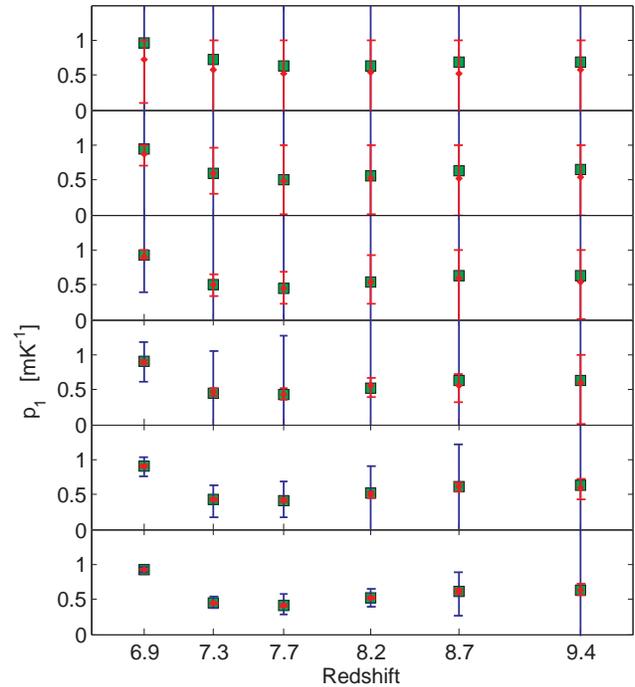}
\caption{Measured value and error of the total probability $p_1$ in the 
first bin ($\Delta T =0-4$ mK) of a 1-bin model of the difference
PDF. We show this as a function of redshift, assuming MWA noise, for
various voxel-pair distance bins: $r_{\rm mid}=$ 4.3, 8.3, 16.2, 31.5,
61.4, and 119.5 comoving Mpc (panels from top to bottom).  We compare
in each case the true value (green square), the mean fit value (red
dot), and 16-84 percentile values (red error bars) based on 1,000
instances of mock observation data; we refer to these percentile
values as the $\pm 1\sigma$ range in the rest of this paper. We also
show the $\pm 1\sigma$ errors magnified by a factor of 10 (blue error
bars), so that smaller errors are easier to see.}
\label{PanelsOfFitvsRedshiftDistanceFigure-sNf1}
\end{figure}

We begin with a 1-bin model consisting of a first bin between $\Delta
T =0$ to 4 mK, plus a normalization bin at $\Delta T = 4$ to 40 mK.
Figure~\ref{PanelsOfFitvsRedshiftDistanceFigure-sNf1} illustrates the
main results for this one-parameter model with one-year MWA noise. 
Even with this relatively large noise, the high number of voxel pairs
in the observational volume of MWA drives the finite sampling noise to
quite low levels, in many cases allowing us to overcome the degeneracy
in the reconstruction. The value of the difference PDF in the first
bin can be measured with an accuracy of a few percent for the larger
voxel-pair distances $r_{\rm mid}=$ 61.4 and 119.5 Mpc, with some
useful measurements possible at lower radii. The measurements are
particularly advantageous at low redshift, where the high ionization
fraction produces a large $\Delta P_D$ feature in the intrinsic
difference PDF (Figure~\ref{TrueSimulatedPDF}); this makes the
measurement easier, and also makes the measured value more closely
related to the ionization fraction. Indeed, the strong rise with time
in the measured 1-bin $p_1$, occurring simultaneously at all $r$ bins,
constitutes a clear detection of the end of reionization (at $z=6.9$
in this case).

\begin{figure}
\centering
\includegraphics[width=1\columnwidth]
{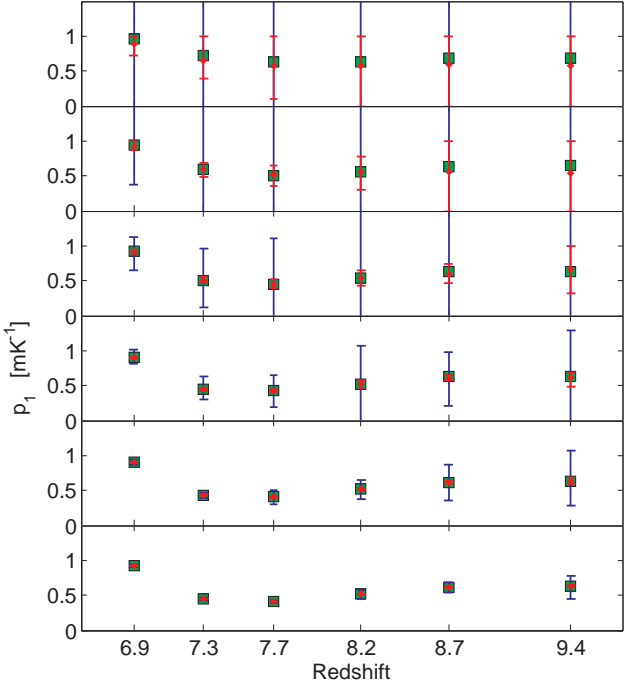}
\caption{Same as Figure~\ref{PanelsOfFitvsRedshiftDistanceFigure-sNf1}, but
for MWA/2 noise, corresponding to 4 years of observations with the
MWA.}
\label{PanelsOfFitvsRedshiftDistanceFigure-sNf2}
\end{figure}

As shown in Figure~\ref{PanelsOfFitvsRedshiftDistanceFigure-sNf2},
decreasing the noise by a factor of 2 (MWA/2) often decreases the
errors in the reconstructed difference PDF by a factor of $3-4$,
yielding some information even at the lowest values of $r$. Since
values of $r \sim 10$ Mpc are required to see the variation of
$p_\Delta$ with distance (Figure~\ref{TrueSimulatedPDF}), this lower
noise would allow us to get an indication of the average size of
ionized bubbles, above which the correlation strength (and $p_1$)
drops.

\begin{figure}
\centering
\includegraphics[width=1\columnwidth]
{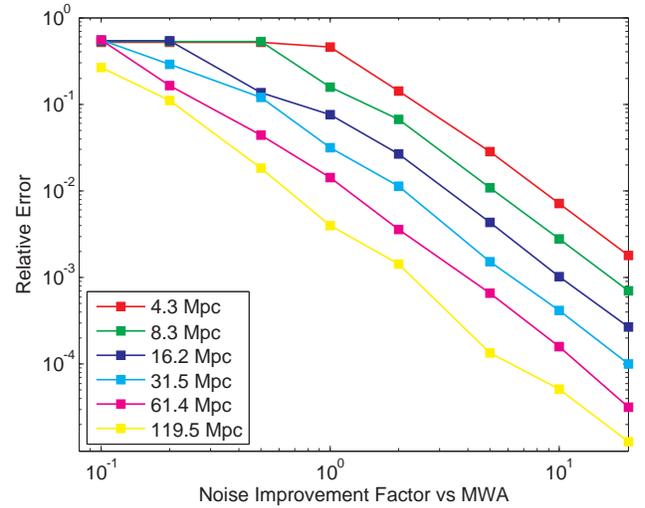}
\caption{Relative error of the measured $p_1$ in the 
first bin ($\Delta T =0-4$ mK) of the 1-bin model as a function of
noise at $z=6.9$, when the mean ionization fraction was
$\overline{x}_i = 0.98$.  Here the noise improvement factor is the
factor by which the thermal noise is reduced compared to 1-year MWA
observations (equation~(\ref{Eq_sigmaN})). The relative error is
defined as the range between the $\pm 1\sigma$ values divided by the
true value.}
\label{ErrorVsNoise-z69}
\end{figure}

\begin{figure}
\centering
\includegraphics[width=1\columnwidth]
{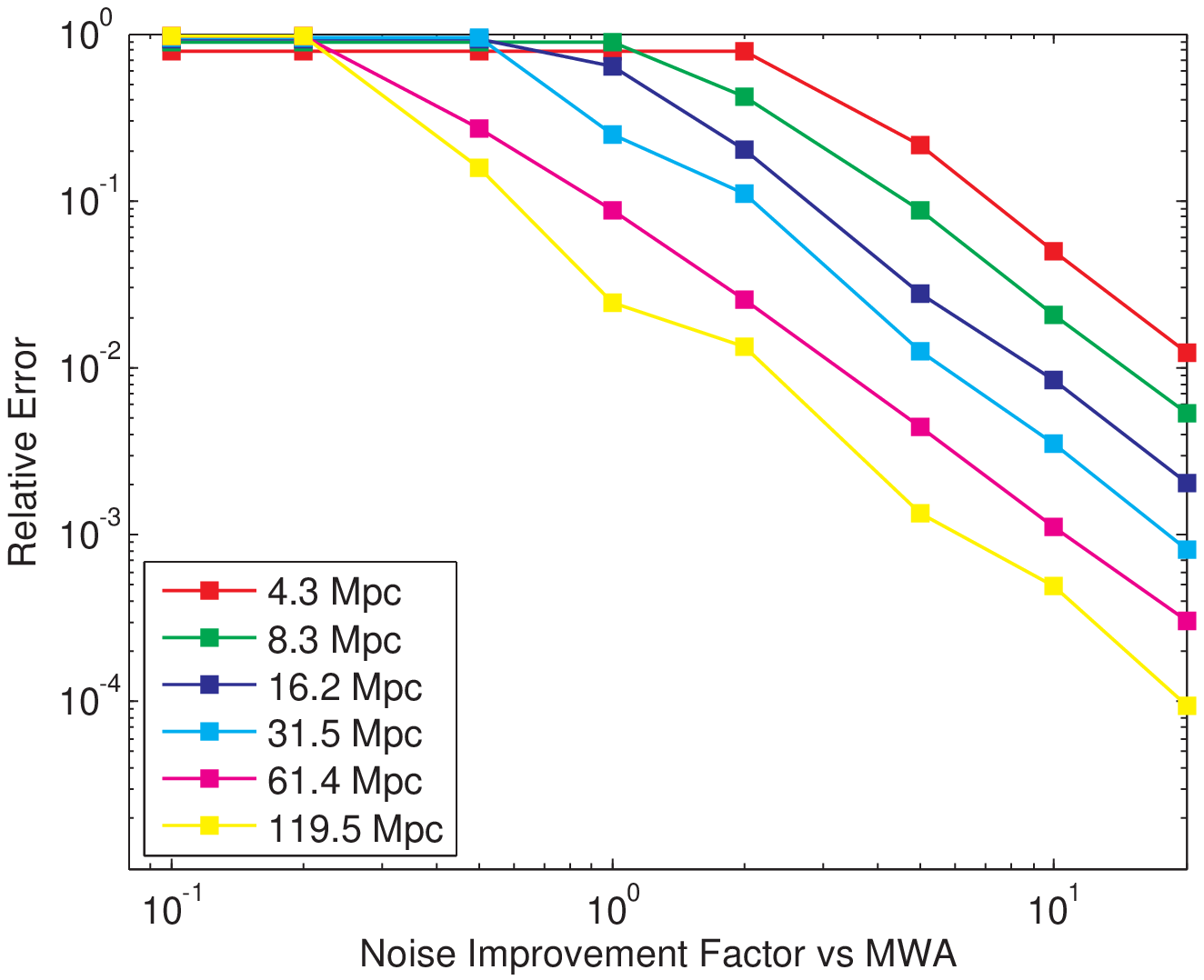}
\caption{Same as Figure~\ref{ErrorVsNoise-z69} but at $z=8.2$, 
when $\overline{x}_i = 0.44$.}
\label{ErrorVsNoise-z82}
\end{figure}

\begin{figure}
\centering
\includegraphics[width=1\columnwidth]
{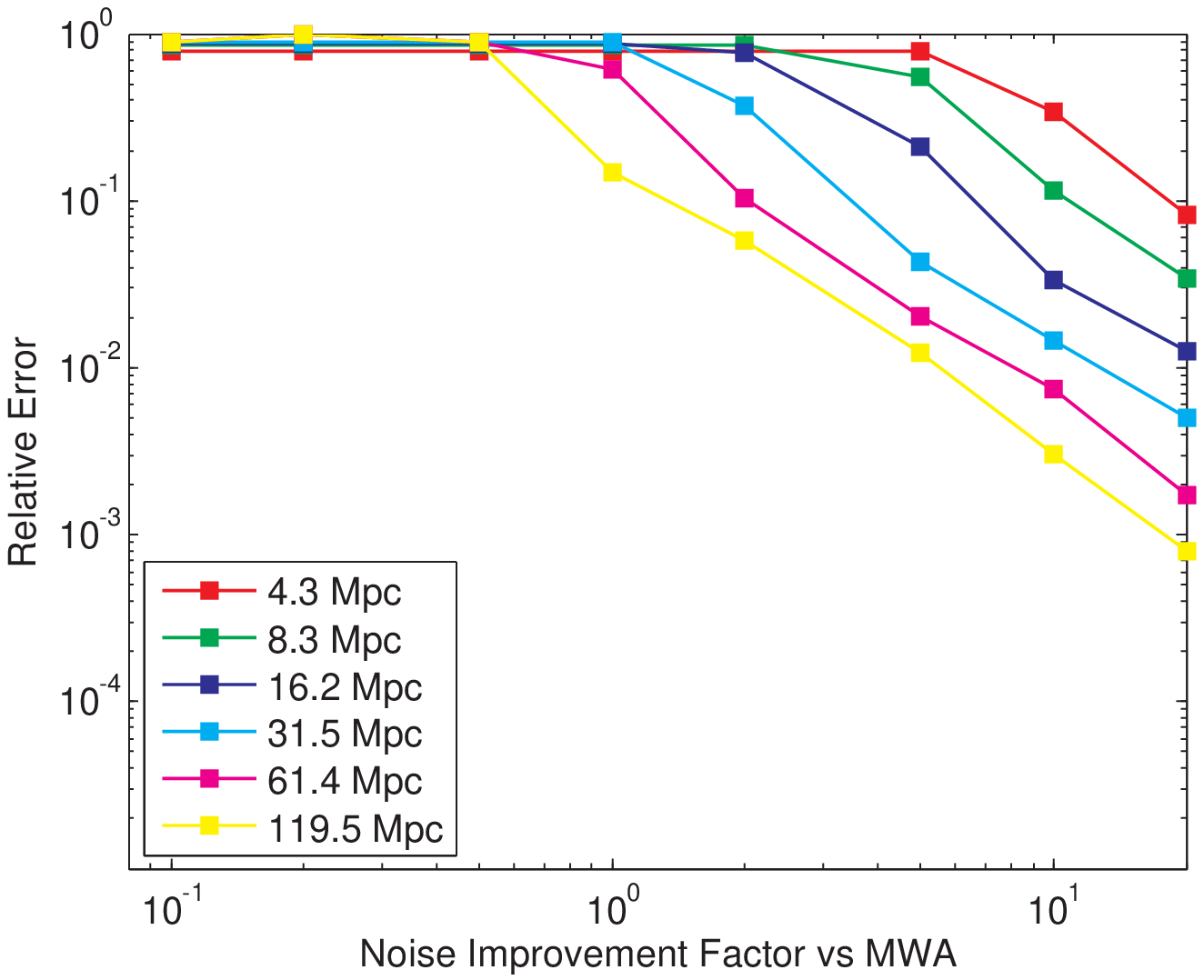}
\caption{Same as Figure~\ref{ErrorVsNoise-z69} but at $z=10.1$, when 
$\overline{x}_i = 0.13$.}
\label{ErrorVsNoise-z101}
\end{figure}

Figures \ref{ErrorVsNoise-z69}, \ref{ErrorVsNoise-z82} and
\ref{ErrorVsNoise-z101} give a more complete 
indication of how the fit error varies with the thermal noise, at the
end, middle, and beginning of the epoch of reionization, respectively.
Again, we find that the larger distance bins have smaller errors
because they have far more voxel-pairs, reducing the sampling error.
Above some high level of noise, the degeneracy is complete and the fit
error is of order unity regardless of the noise per pixel $\sigma_N$
(note that the probability within a bin is limited to vary between 0
and 1). However, below some critical value (which varies with $r$ due
to the different numbers of voxel-pairs), the fit error begins to
decrease as $\sigma_N$ decreases. This decrease is faster than linear
(typically close to quadratic) since reduced noise removes some of the
partial degeneracy involved in the effective deconvolution of
$p_\Delta$. Note that the low noise levels we consider can correspond
either to multi-year observations with the MWA or to future radio
arrays with a larger collecting area. 

At $z=6.9$, the MWA suffices to measure the 1-bin difference PDF with
$<10\%$ errors down to the $r=16$ Mpc bin, and thus verify the
signature of the end of reionization. At the larger separations, the
MWA can determine $p_\Delta$ to better than a percent. Higher
redshifts are more challenging, so that at $z=8.2$ (near the midpoint
of reionization), $<10\%$ errors are possible with the MWA only at the
two highest separations, MWA/2 gets down to $\sim 30$ Mpc, and MWA/10
allows measurements at the full range of separations. Early in
reionization ($z=10.1$), when the measurement noise is larger due to
the higher redshift, and the difference PDF is still close to the
Gaussian shape driven by density fluctuations, the MWA can only
attempt to measure the long-separation limit of $p_\Delta$, but a
second-generation experiment should still be able to probe a broad
range of distances.

\begin{figure}
\centering
\includegraphics[width=1\columnwidth]
{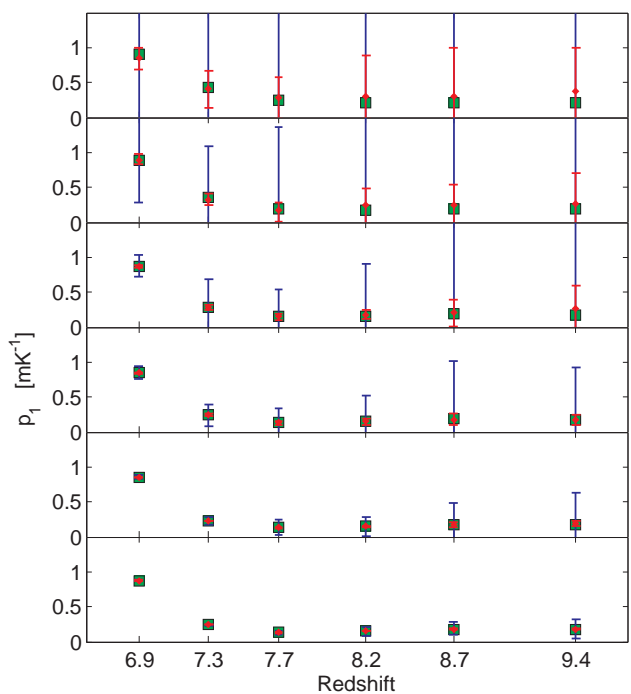}
\caption{Measured value and error of the total probability $p_1$ in the 
first bin of a 1-bin model of the difference PDF. Same setup and
notation as in Figure~\ref{PanelsOfFitvsRedshiftDistanceFigure-sNf1},
but with MWA/2 noise as in
Figure~\ref{PanelsOfFitvsRedshiftDistanceFigure-sNf2}, and here using
a model with a narrower first bin ($\Delta T =0-1$ mK).}
\label{PanelsOfFitvsRedshiftDistanceFigure-BinF1-sNf2}
\end{figure}

\begin{figure}
\centering
\includegraphics[width=1\columnwidth]
{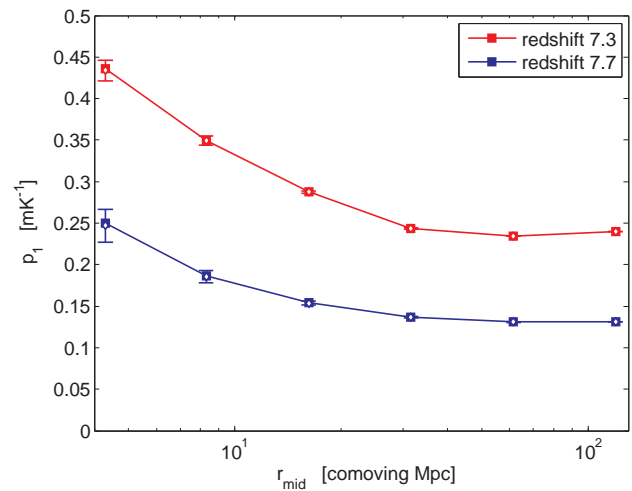}
\caption{Measured value and error of the total probability $p_1$ in the 
first bin ($\Delta T =0\sim 1$ mK) of a narrow 1-bin model of
$p_\Delta$, shown as a function of voxel-pair distance (bin center),
with MWA/10 noise.}
\label{Fit_vs_rmid-BinF1-sNf10}
\end{figure}

Alternatively, we can use a 1-bin model where the first bin is
smaller, between $\Delta T =0$ to 1 mK. This pinpoints the fraction of
voxel-pairs with $\Delta T_b\approx 0$ (which approximates $\Delta
P_D$) more accurately. Thus, $p_1$ is significantly lower than with
the wider first bin consider before, except at the very end of
reionization. The measurement errors are generally similar to before,
as illustrated for MWA/2 noise in
Figure~\ref{PanelsOfFitvsRedshiftDistanceFigure-BinF1-sNf2}, which can
be compared to Figure~\ref{PanelsOfFitvsRedshiftDistanceFigure-sNf2}
from before. Figure~\ref{Fit_vs_rmid-BinF1-sNf10} uses a different
presentation to show more directly how with second-generation radio
arrays we can accurately map $\Delta P_D$ as a function of $r$ across
different redshifts, with the location of the flat asymptote of $p_1$
telling us where $r$ drops below the correlation length, which can be
used to determine the average size of ionized bubbles.  We note that
one conservative way to begin investigating the shape of the
difference PDF is to fit 1-bin models with various bin widths and
compare the results. However, a more direct method is to use models
with more bins, which we consider next.

\subsection{10-bin model}

Given that the 1-bin model is expected to yield accurate measurements
of the difference PDF, even better than $1\%$ measurements in some
cases, we now consider a more ambitious attempt to measure the
detailed shape of the difference PDF. We consider a 10-bin model
consisting of 10 equal-size bins between $\Delta T =0$ and $10$ mK,
plus a normalization bin at $\Delta T=10$ to 40 mK. Of course, many
other binning choices are possible, including redshift-dependent
binning, but our choice should suffice to determine whether the shape
of the PDF can be determined with 1~mK bins over a range where there
is interesting dependence on $\Delta T$ throughout the reionization
era. 

\begin{figure*}
\centering
\subfloat[Subfigure 1 list of figures text][MWA]{
\includegraphics[width=0.31\textwidth]
{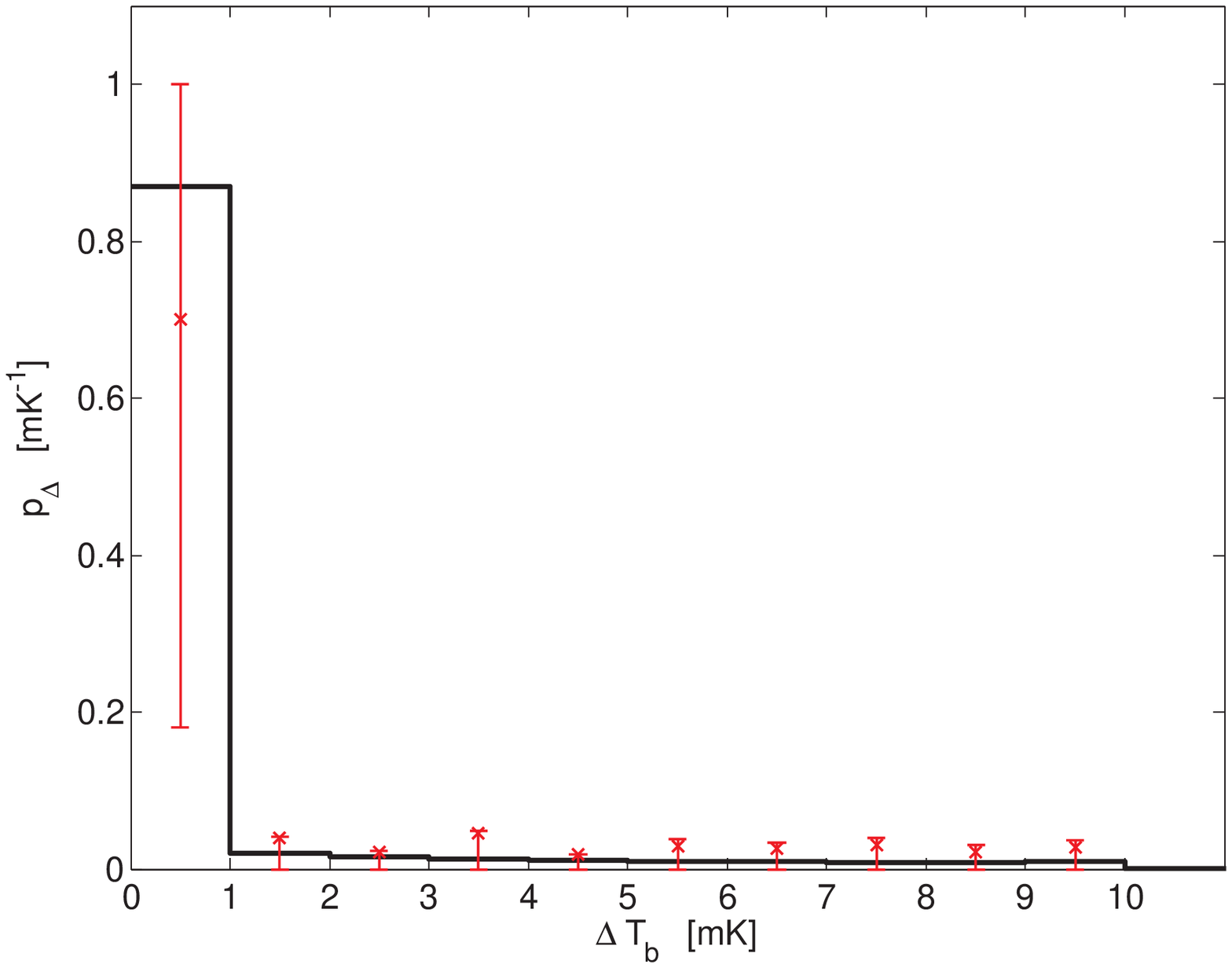}
\label{stair_figure-z69-rI3:SNf1}}
\quad
\subfloat[Subfigure 1 list of figures text][MWA/2]{
\includegraphics[width=0.31\textwidth]
{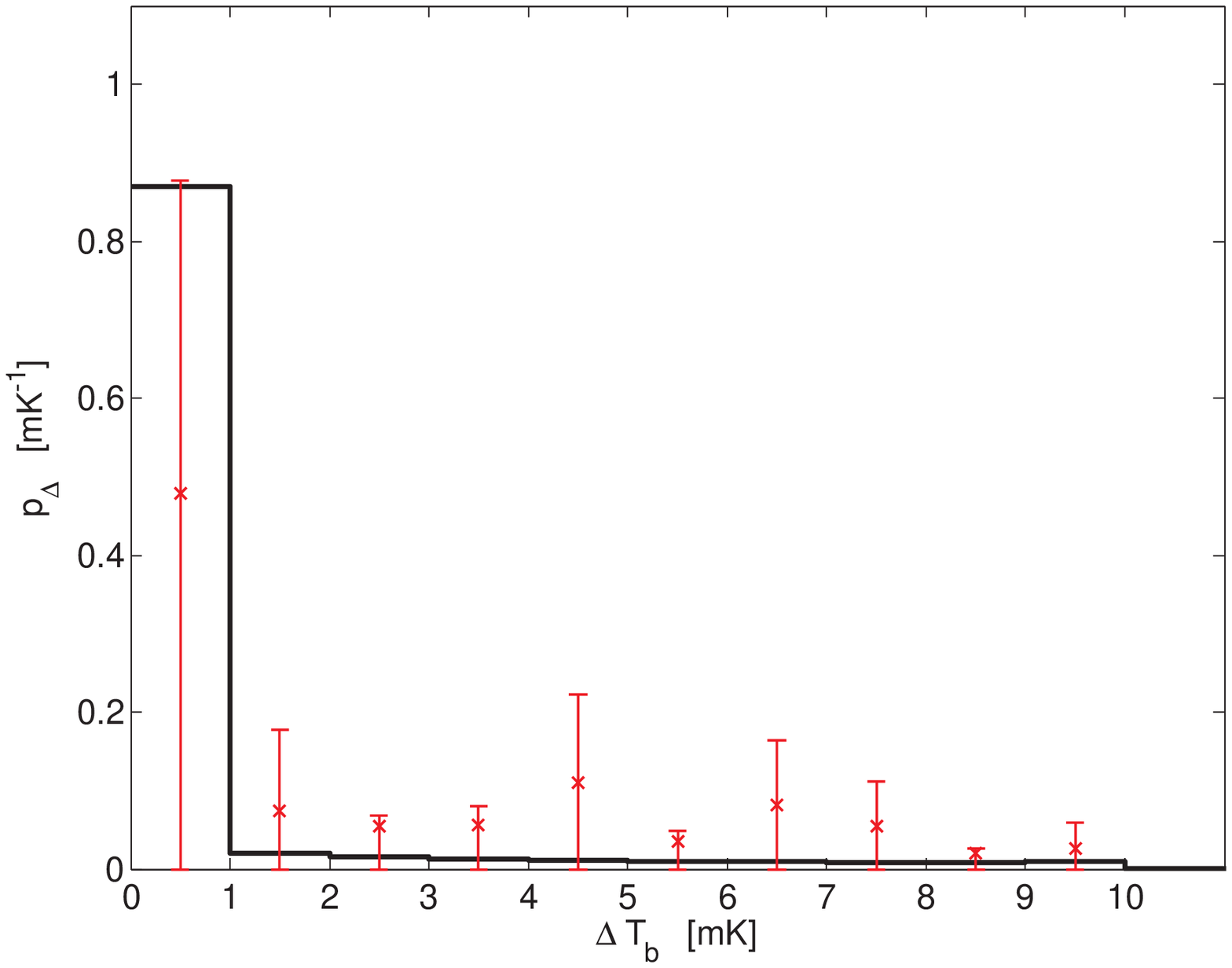}
\label{stair_figure-z69-rI3:SNf2}}
\quad
\subfloat[Subfigure 2 list of figures text][MWA/10]{
\includegraphics[width=0.31\textwidth]
{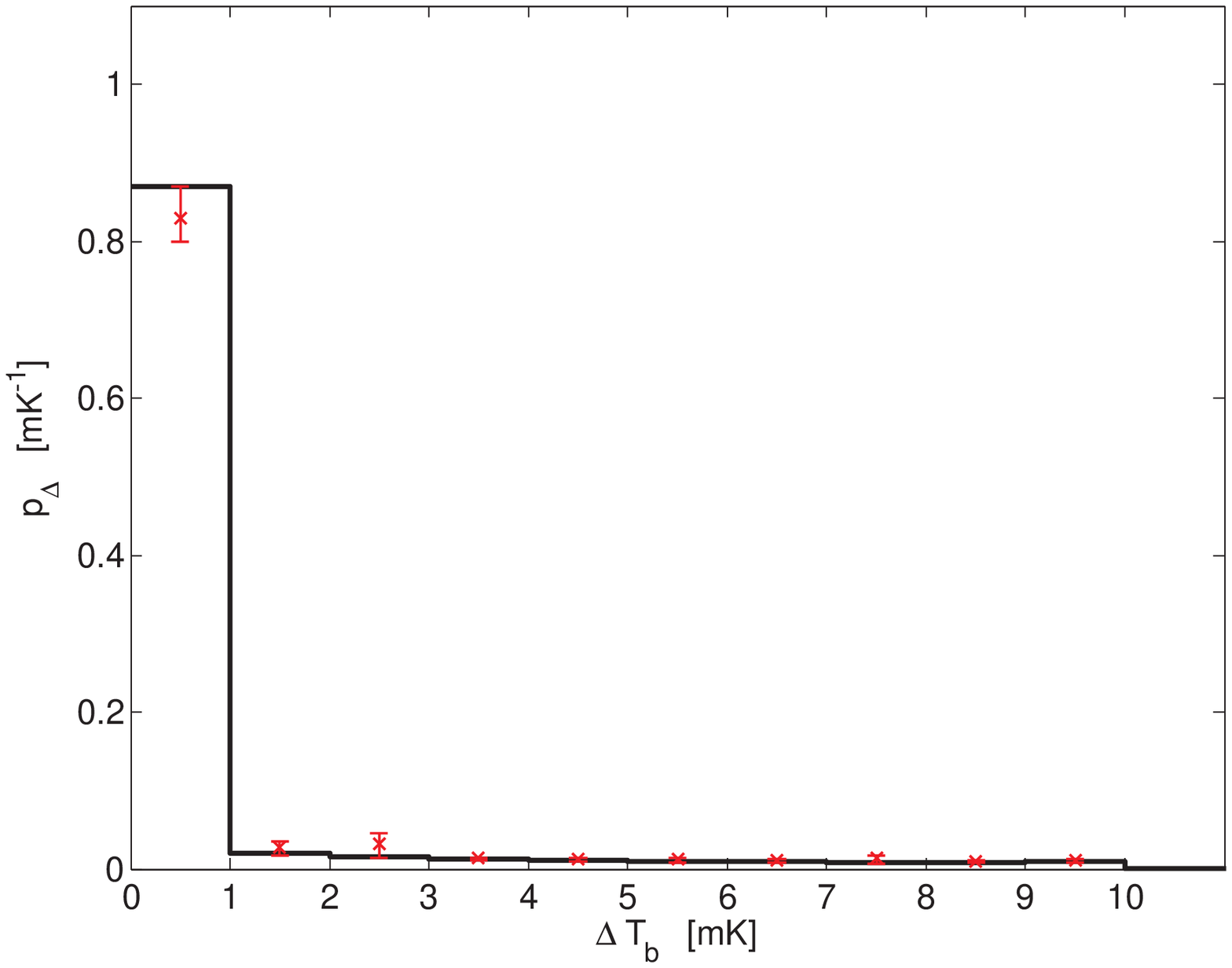}
\label{stair_figure-z69-rI3:SNf10}}
\caption{Measured value and $1-\sigma$ error of $p_\Delta$ in the 10-bin 
model, at $z=6.9$, in the $r_{\rm mid}=16.2$ Mpc bin, shown for three
different levels of the thermal noise (as indicated in each panel). We
compare the true, input difference PDF (black line) in 10 bins of
width 1~mK (an extra normalization bin for at $\Delta T>10$ mK is not
shown) to the mean value and $16-84$ percentile range of the
reconstructed $p_\Delta$ based on fitting to 1,000 mock data sets.}
\label{stair_figure-z69-rI3:globfig}
\end{figure*}

\begin{figure*}
\centering
\subfloat[Subfigure 1 list of figures text][MWA]{
\includegraphics[width=0.31\textwidth]
{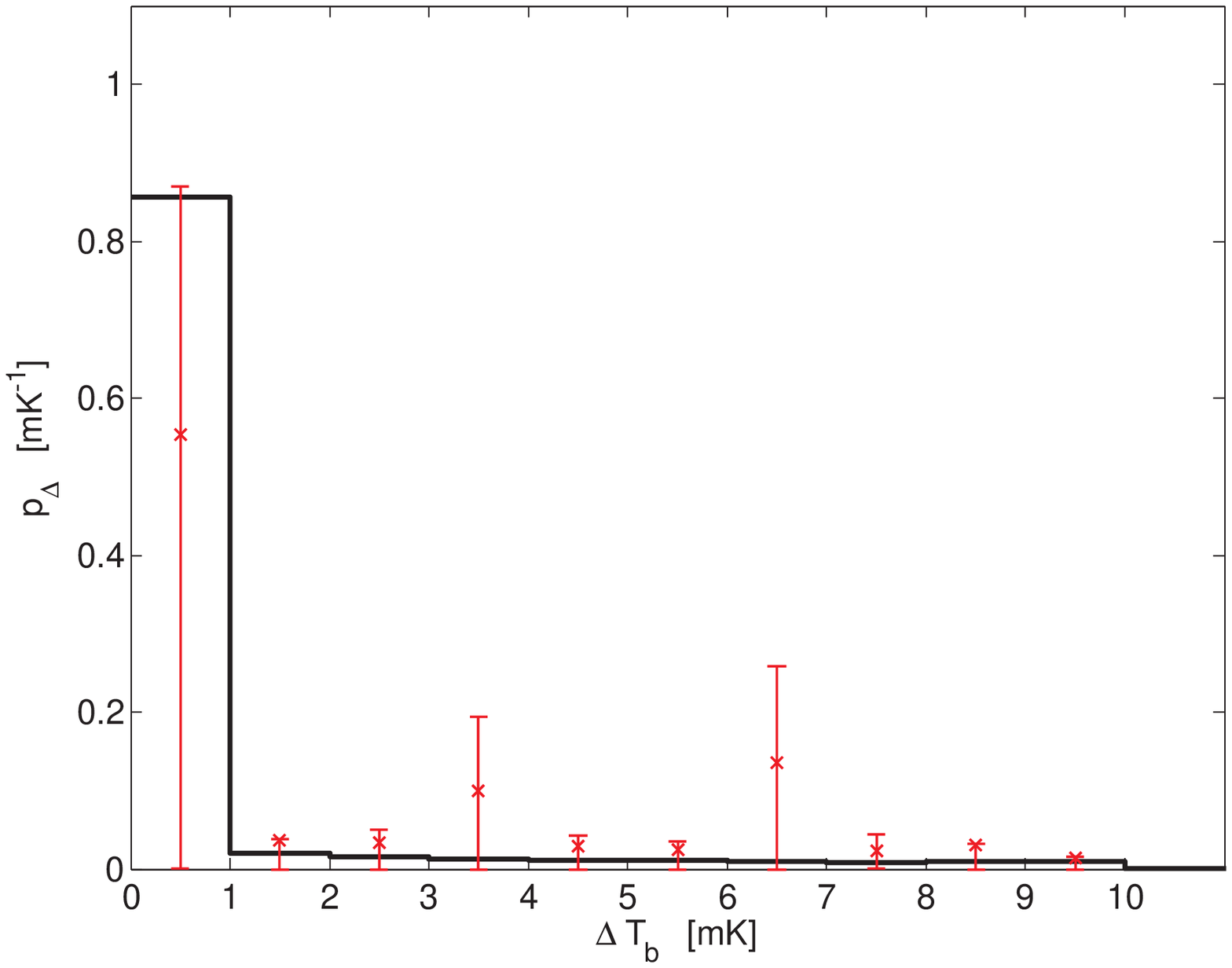}
\label{stair_figure-z69-rI5:SNf1}}
\quad
\subfloat[Subfigure 1 list of figures text][MWA/2]{
\includegraphics[width=0.31\textwidth]
{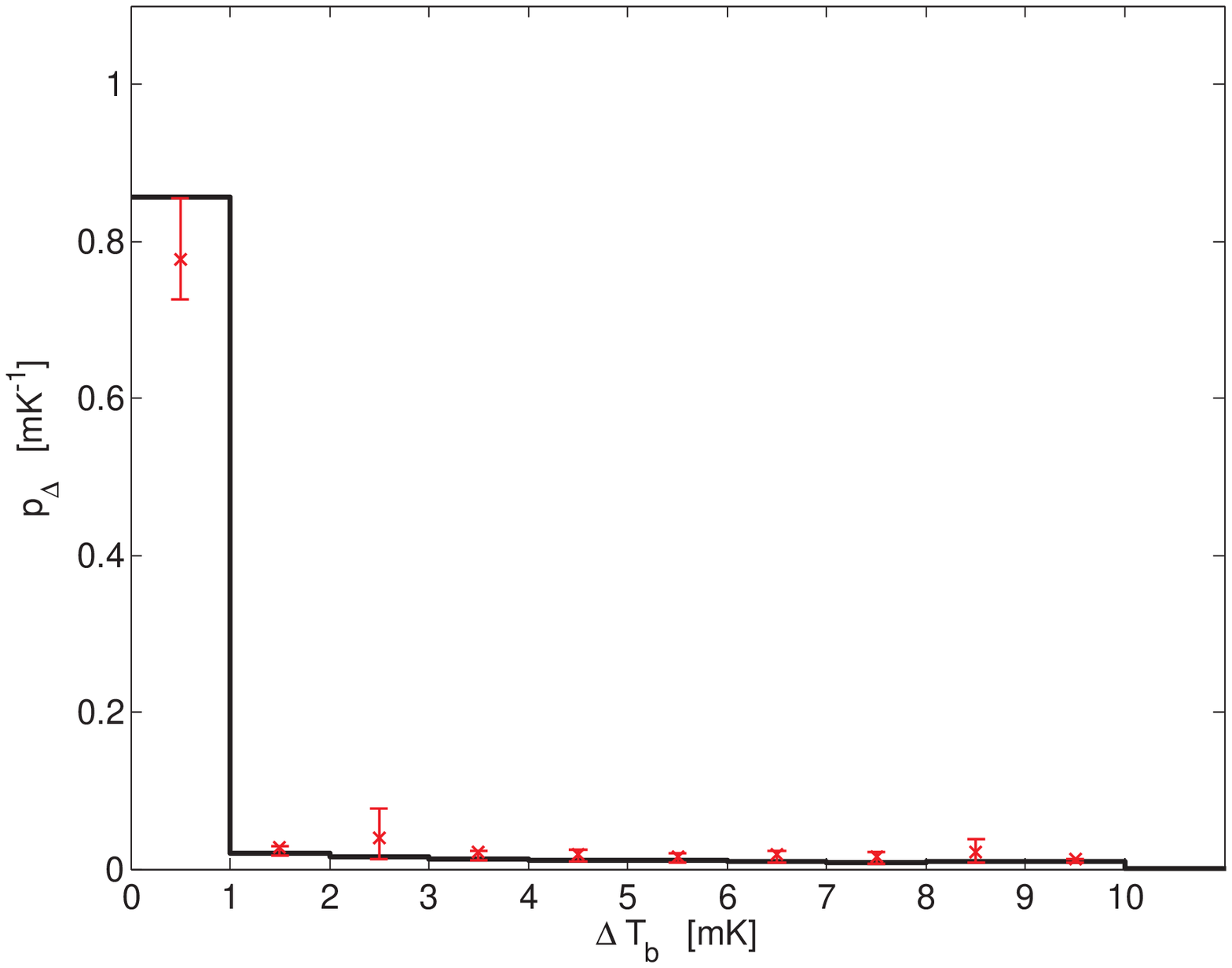}
\label{stair_figure-z69-rI5:SNf2}}
\quad
\subfloat[Subfigure 2 list of figures text][MWA/10]{
\includegraphics[width=0.31\textwidth]
{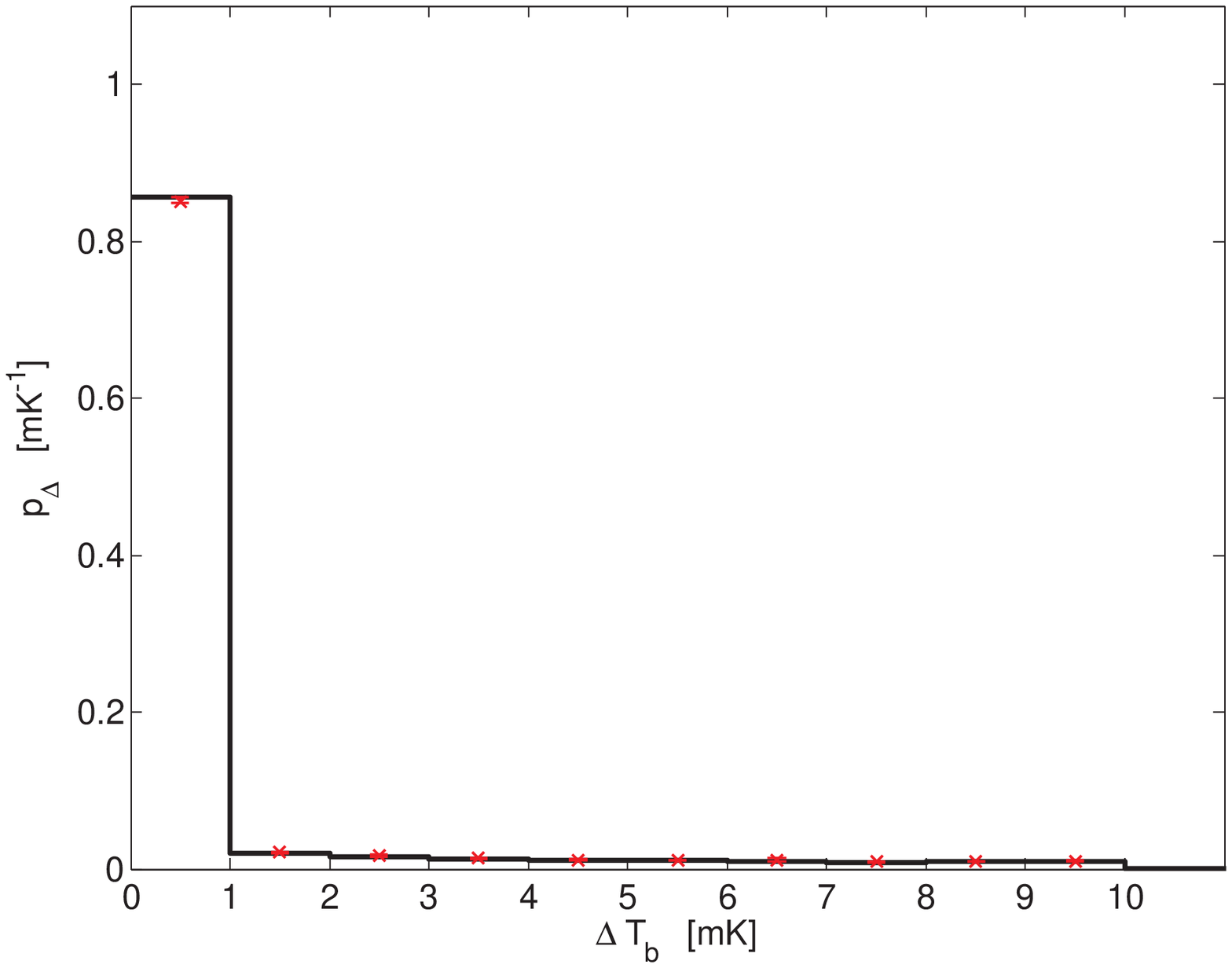}
\label{stair_figure-z69-rI5:SNf10}}
\caption{Same as Figure~\ref{stair_figure-z69-rI3:globfig} but for the
$r_{\rm mid}=61.4$ Mpc bin.}
\label{stair_figure-z69-rI5:globfig}
\end{figure*}

\begin{figure*}
\centering
\subfloat[Subfigure 1 list of figures text][MWA]{
\includegraphics[width=0.31\textwidth]
{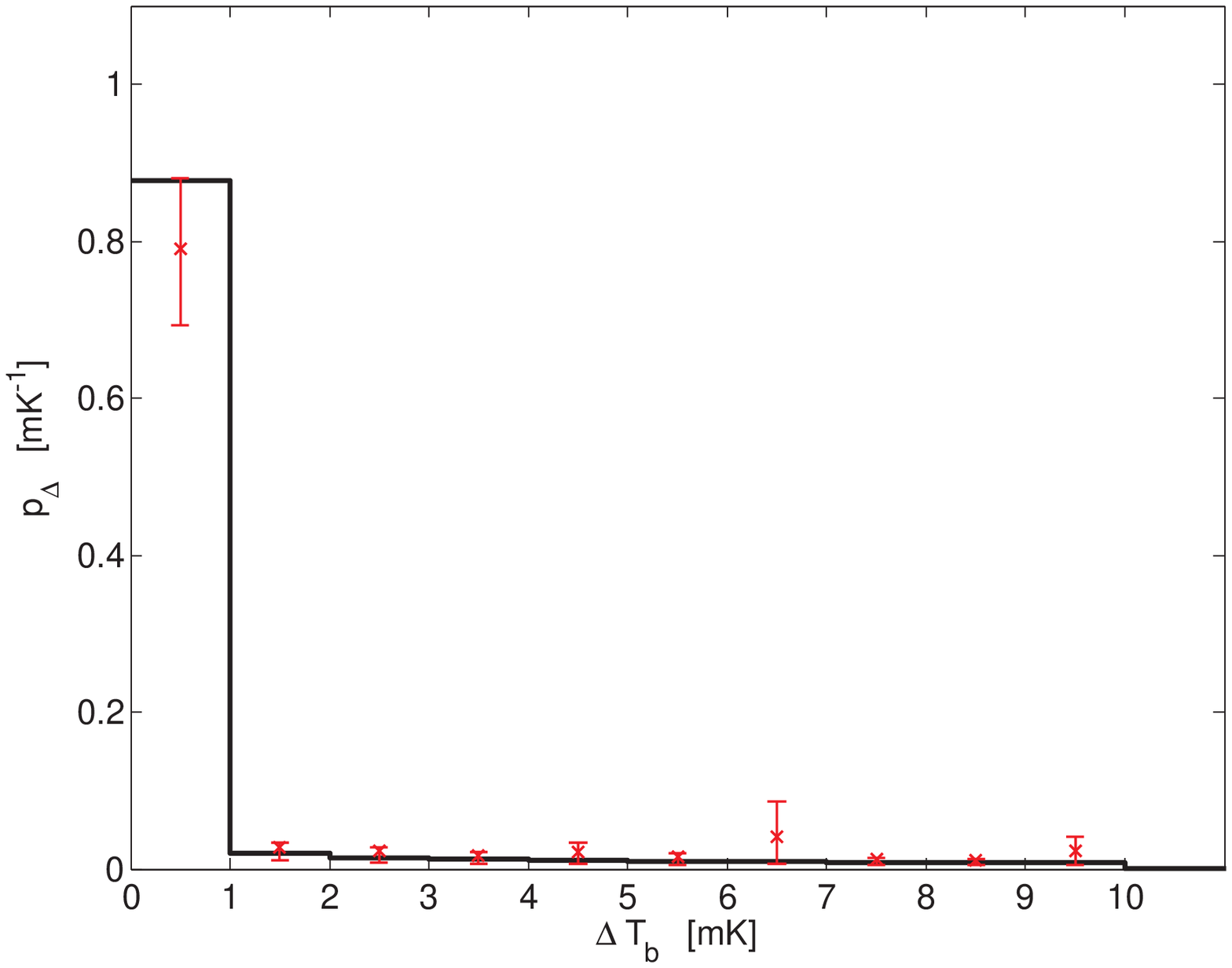}
\label{stair_figure-z69-rI6:SNf1}}
\quad
\subfloat[Subfigure 1 list of figures text][MWA/2]{
\includegraphics[width=0.31\textwidth]
{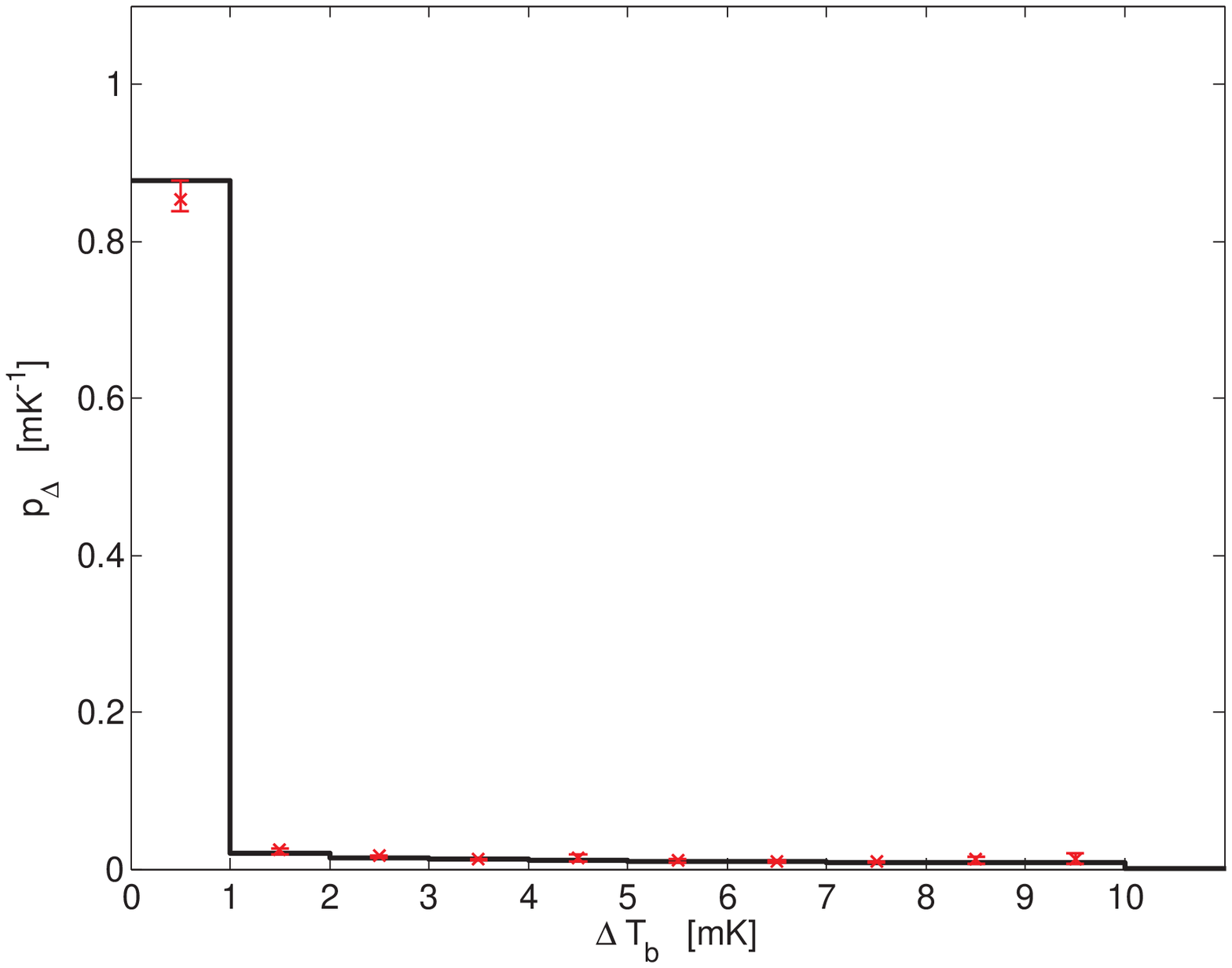}
\label{stair_figure-z69-rI6:SNf2}}
\quad
\subfloat[Subfigure 2 list of figures text][MWA/10]{
\includegraphics[width=0.31\textwidth]
{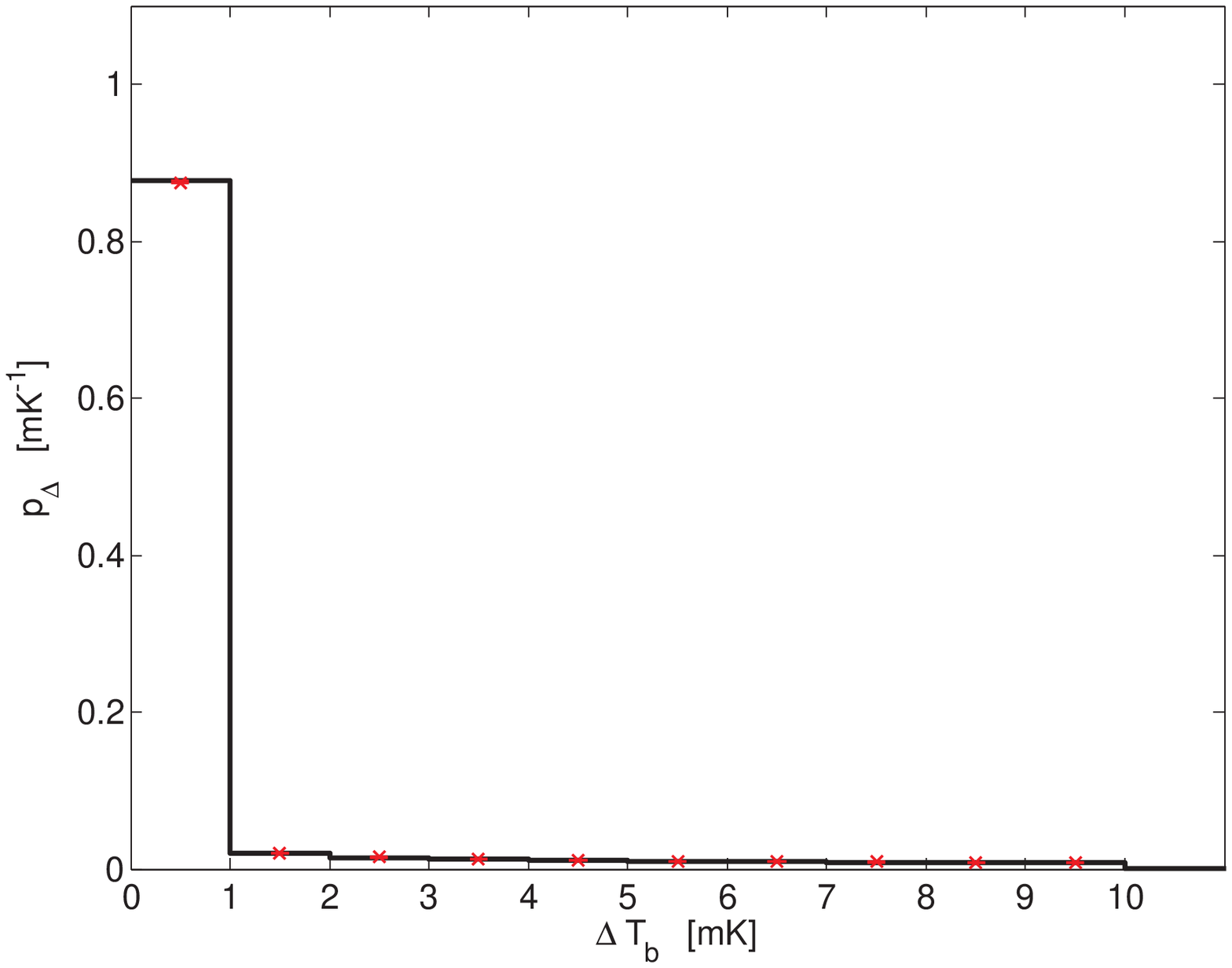}
\label{stair_figure-z69-rI6:SNf10}}
\caption{Same as Figure~\ref{stair_figure-z69-rI3:globfig} but for the
$r_{\rm mid}=119.5$ Mpc bin.}
\label{stair_figure-z69-rI6:globfig}
\end{figure*}

Near the end of reionization, we illustrate the expected
reconstruction accuracy in three separation bins, $r_{\rm mid}=$ 16.2
(Figure~\ref{stair_figure-z69-rI3:globfig}), 61.4
(Figure~\ref{stair_figure-z69-rI5:globfig}) and 119.5 Mpc
(Figure~\ref{stair_figure-z69-rI6:globfig}). While the theoretical
difference PDF should have a simple shape, with nearly all the
probability concentrated in the first bin, it would be exciting to
directly verify this observationally. At 16.2 Mpc, the error in the
first bin is very large, and there are strong degeneracies among the
various bins, as illustrated by the failure of the fitting errors to
decrease in going from MWA to MWA/2 errors. The degeneracy is broken,
however, with MWA/10 errors, in which case the expected shape of
$p_\Delta$ can be precisely verified. We note that the $r_{\rm mid}=$
8.3 Mpc bin (not shown) shows similar reconstruction errors to the
16.2 Mpc case (except that the errors are larger by a factor of a few
for MWA/10). At the two largest-separation bins, $p_\Delta$ can
be reasonably measured with MWA/2 (for $r_{\rm mid}=$ 61.4 Mpc) or 
even with MWA errors (for the highest $r$ bin). 

\begin{figure*}
\centering
\subfloat[Subfigure 1 list of figures text][MWA]{
\includegraphics[width=0.31\textwidth]
{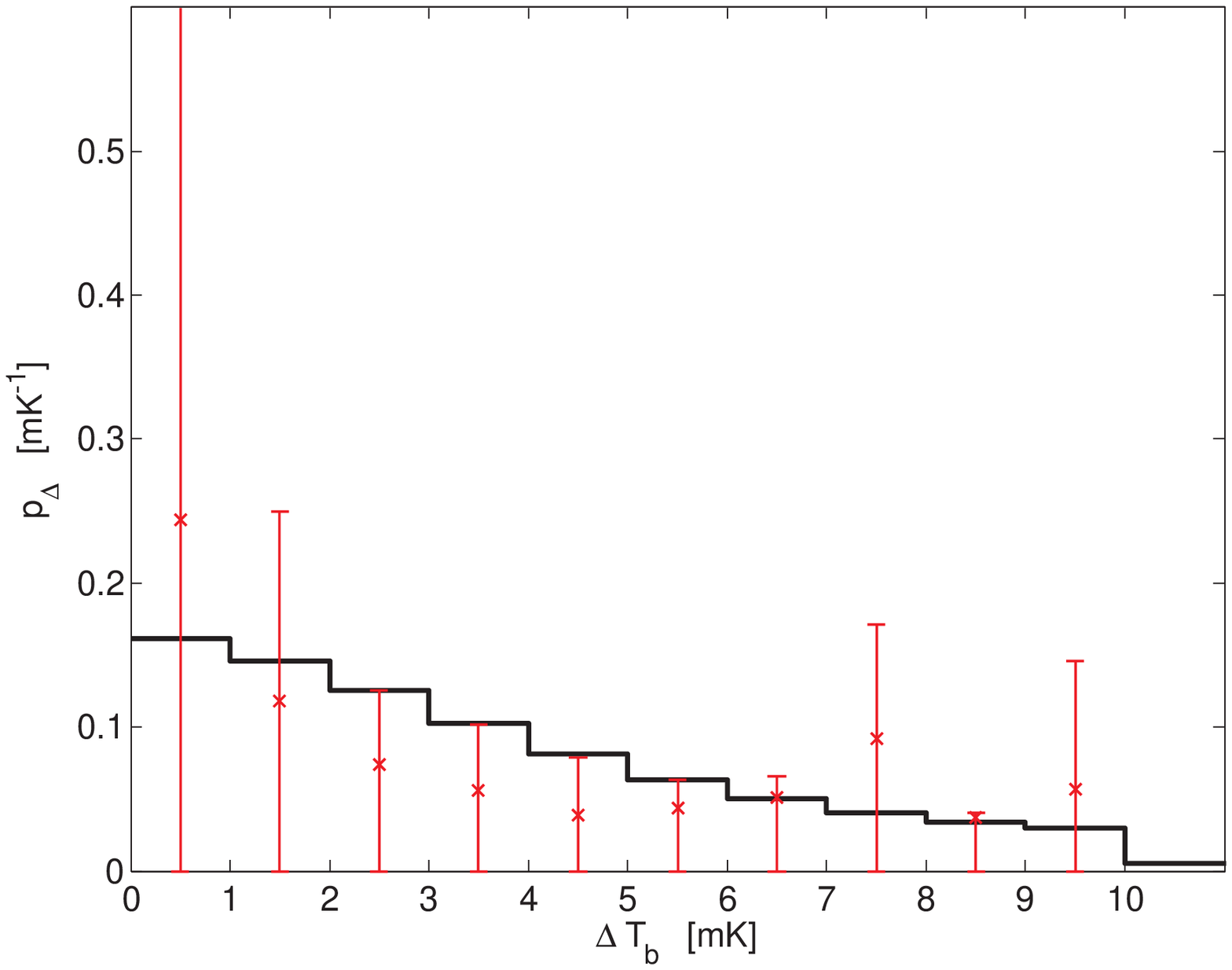}
\label{stair_figure-z82-rI3:SNf1}}
\quad
\subfloat[Subfigure 1 list of figures text][MWA/2]{
\includegraphics[width=0.31\textwidth]
{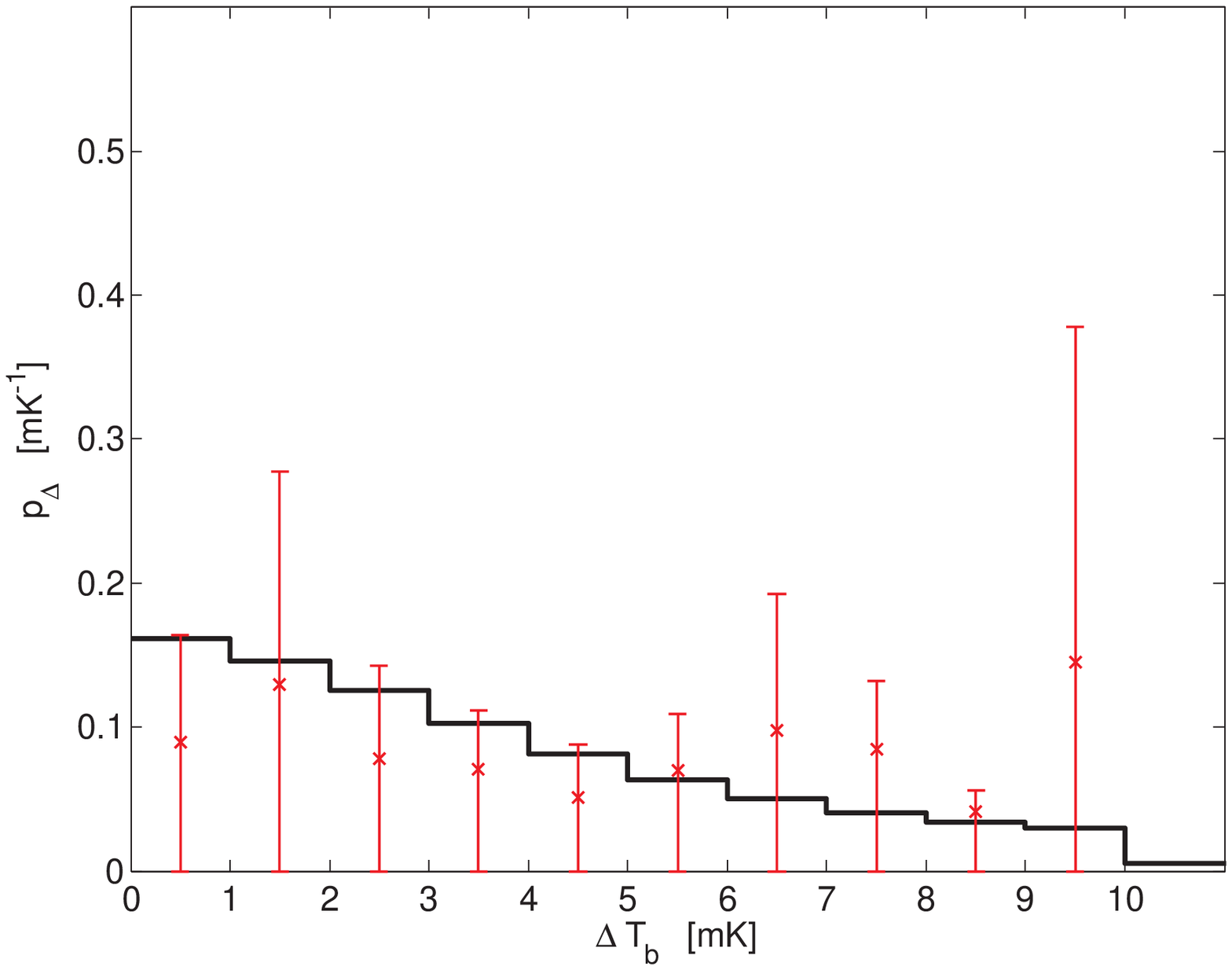}
\label{stair_figure-z82-rI3:SNf2}}
\quad
\subfloat[Subfigure 2 list of figures text][MWA/10]{
\includegraphics[width=0.31\textwidth]
{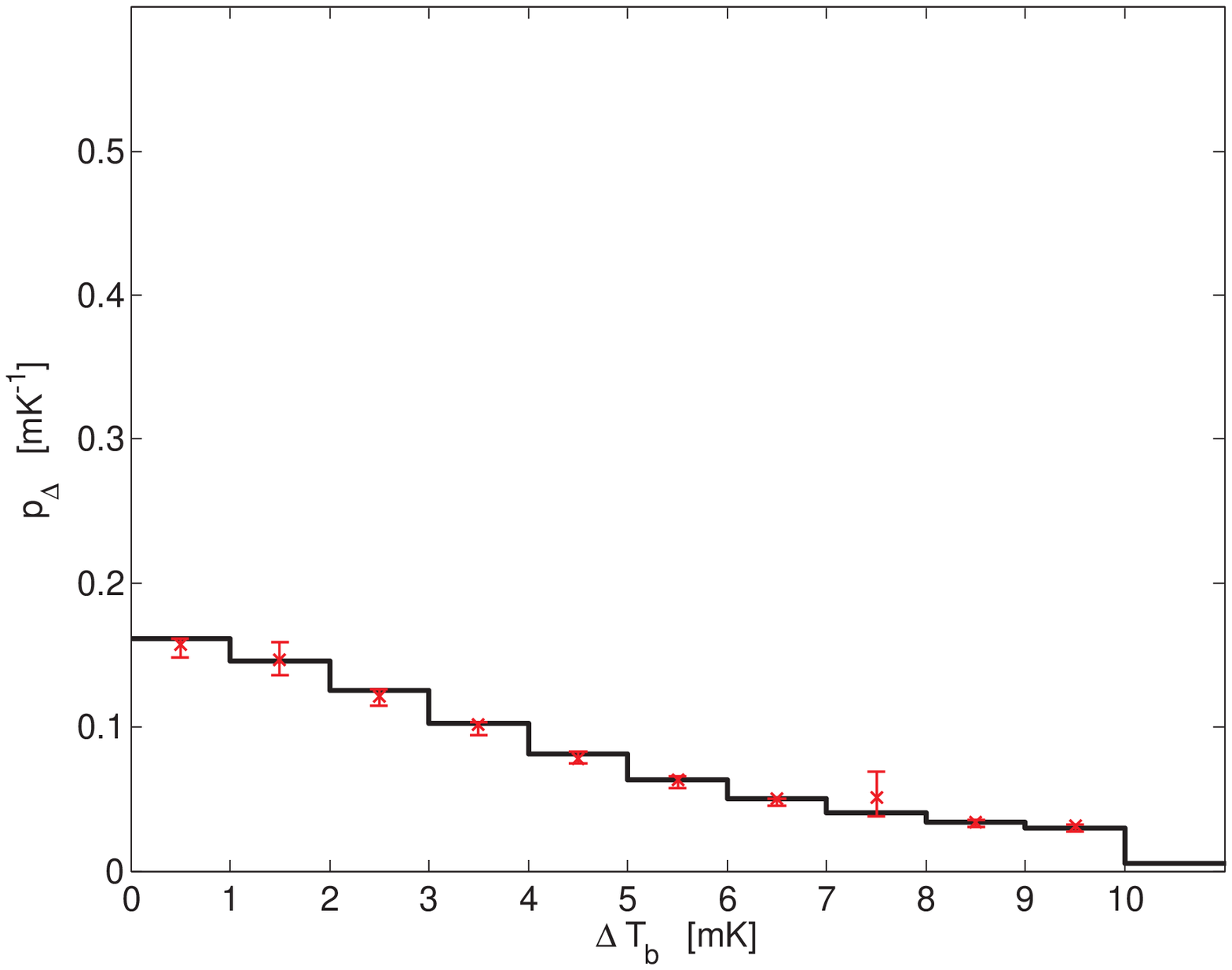}
\label{stair_figure-z82-rI3:SNf10}}
\caption{Same as Figure~\ref{stair_figure-z69-rI3:globfig} (i.e., 
$r_{\rm mid}=16.2$ Mpc) but at $z=8.2$.}
\label{stair_figure-z82-rI3:globfig}
\end{figure*}

\begin{figure*}
\centering
\subfloat[Subfigure 1 list of figures text][MWA]{
\includegraphics[width=0.31\textwidth]
{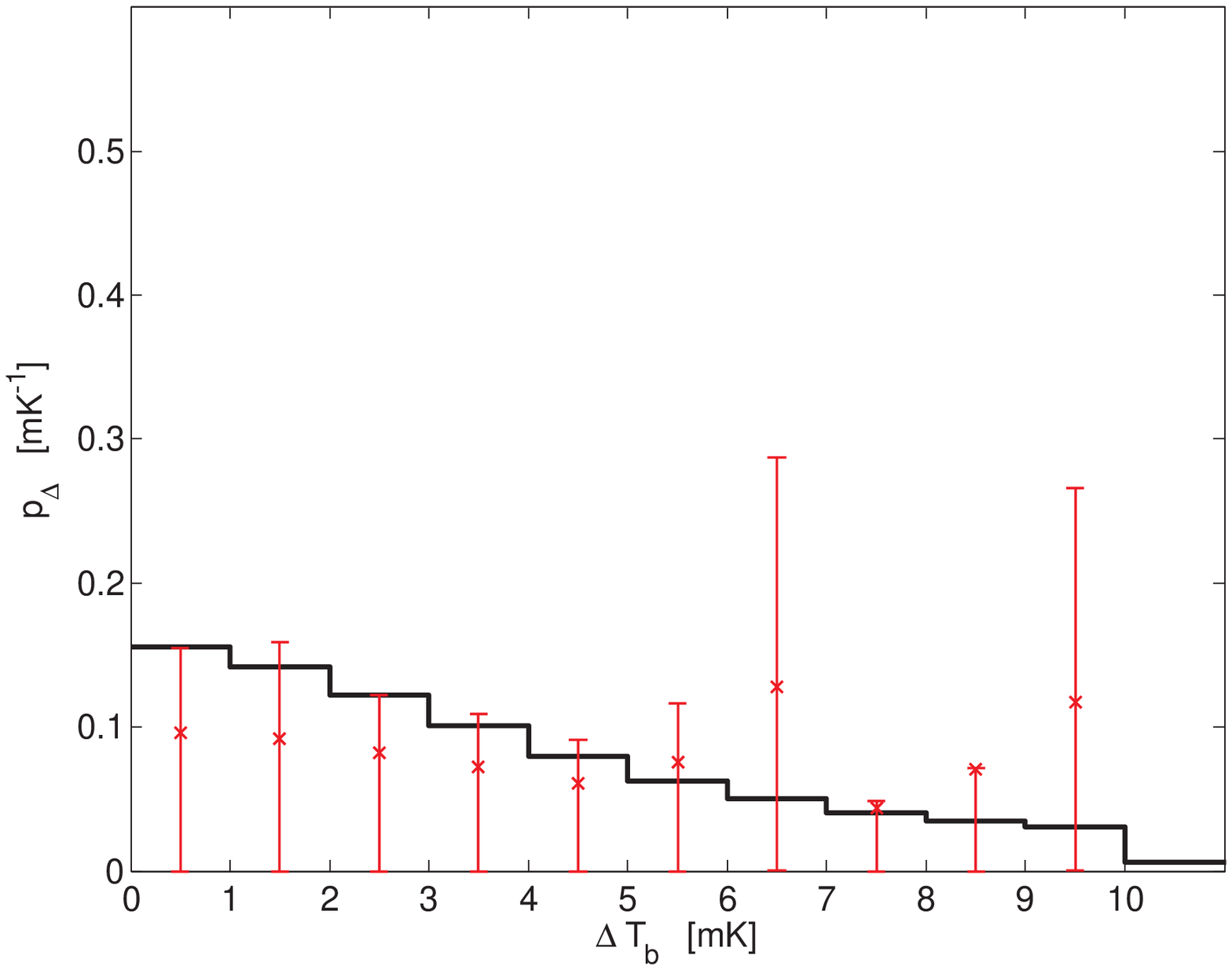}
\label{stair_figure-z82-rI5:SNf1}}
\quad
\subfloat[Subfigure 1 list of figures text][MWA/2]{
\includegraphics[width=0.31\textwidth]
{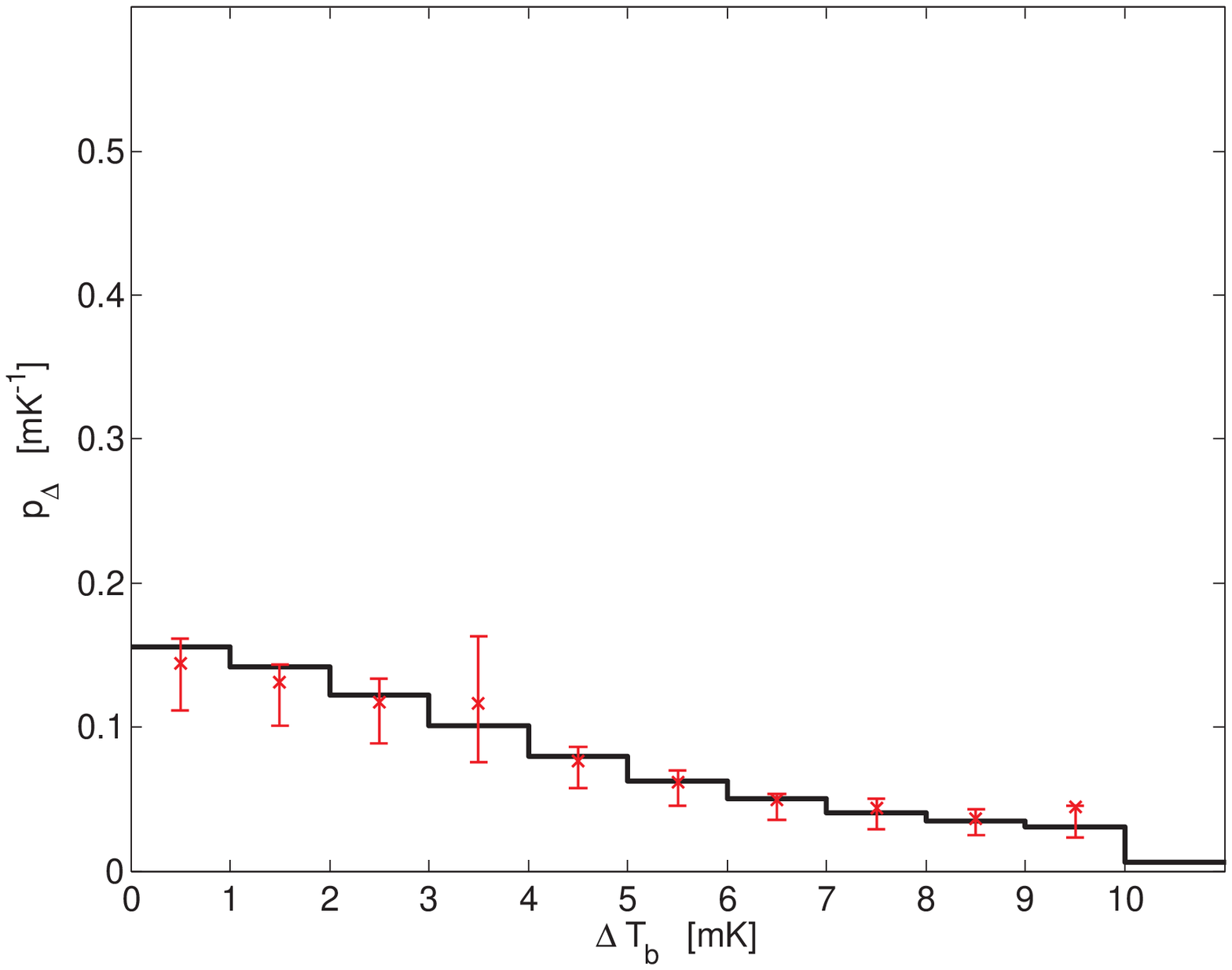}
\label{stair_figure-z82-rI5:SNf2}}
\quad
\subfloat[Subfigure 2 list of figures text][MWA/10]{
\includegraphics[width=0.31\textwidth]
{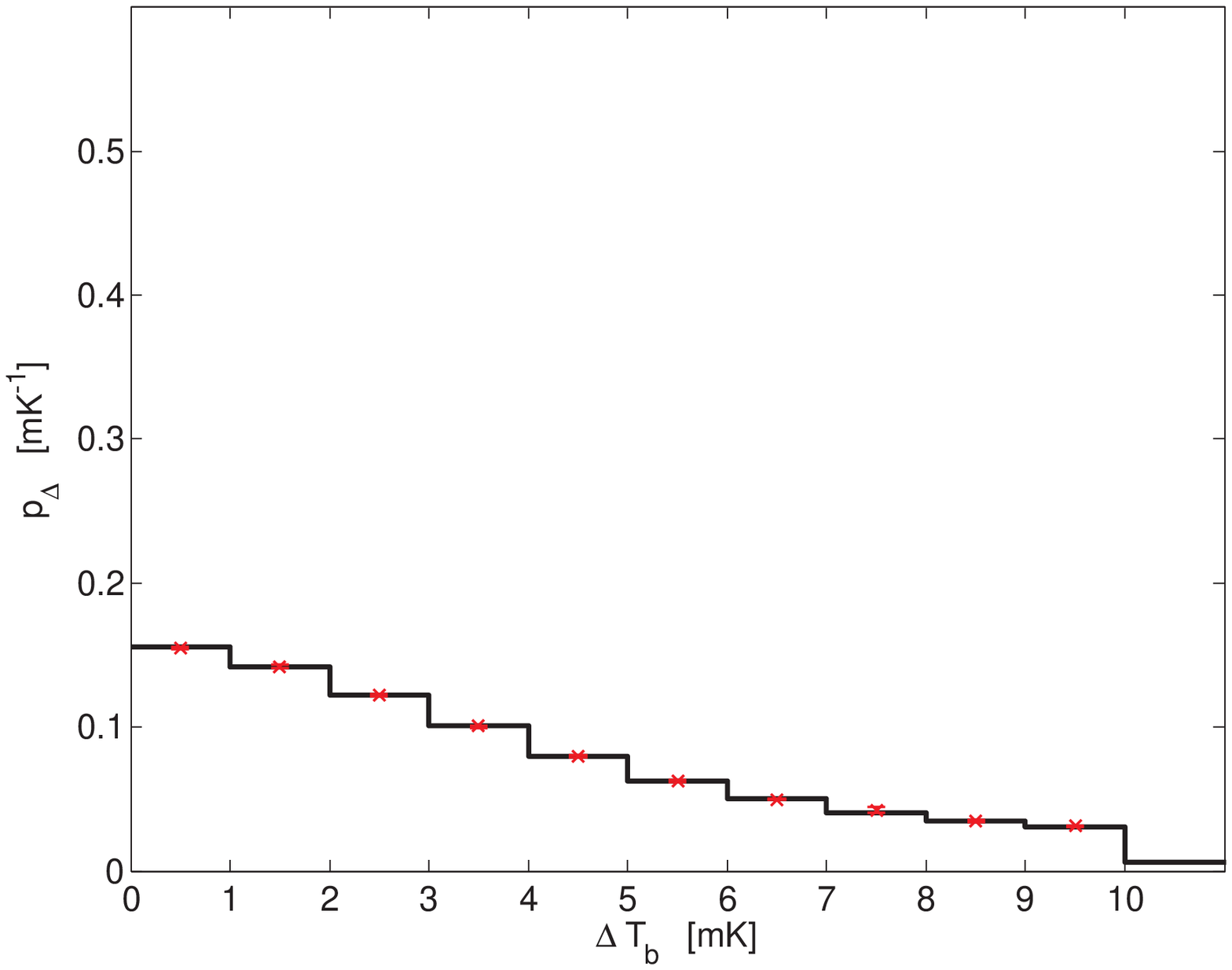}
\label{stair_figure-z82-rI5:SNf10}}
\caption{Same as Figure~\ref{stair_figure-z82-rI3:globfig} but for the
$r_{\rm mid}=61.4$ Mpc bin.}
\label{stair_figure-z82-rI5:globfig}
\end{figure*}

\begin{figure*}
\centering
\subfloat[Subfigure 1 list of figures text][MWA]{
\includegraphics[width=0.31\textwidth]
{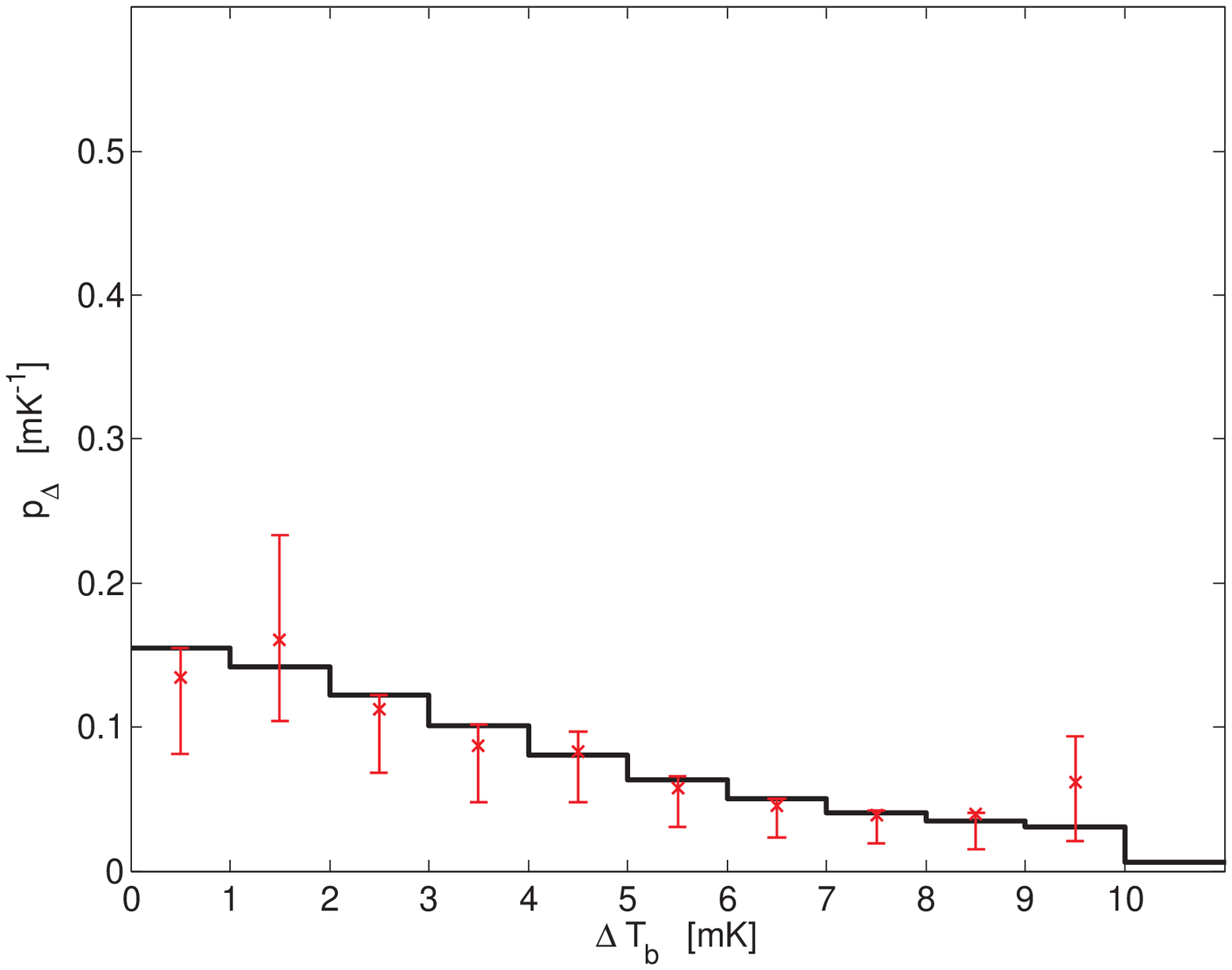}
\label{stair_figure-z82-rI6:SNf1}}
\quad
\subfloat[Subfigure 1 list of figures text][MWA/2]{
\includegraphics[width=0.31\textwidth]
{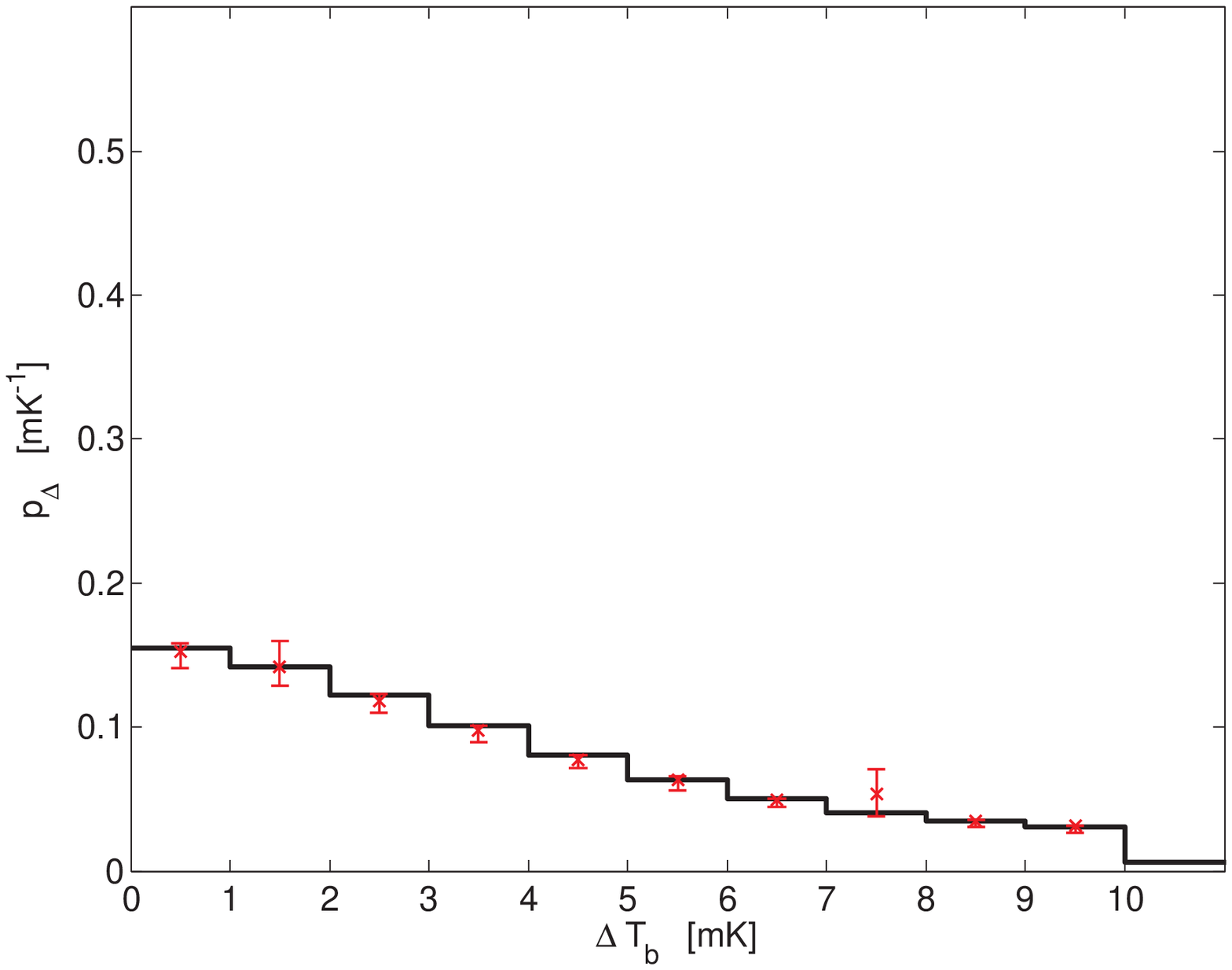}
\label{stair_figure-z82-rI6:SNf2}}
\quad
\subfloat[Subfigure 2 list of figures text][MWA/10]{
\includegraphics[width=0.31\textwidth]
{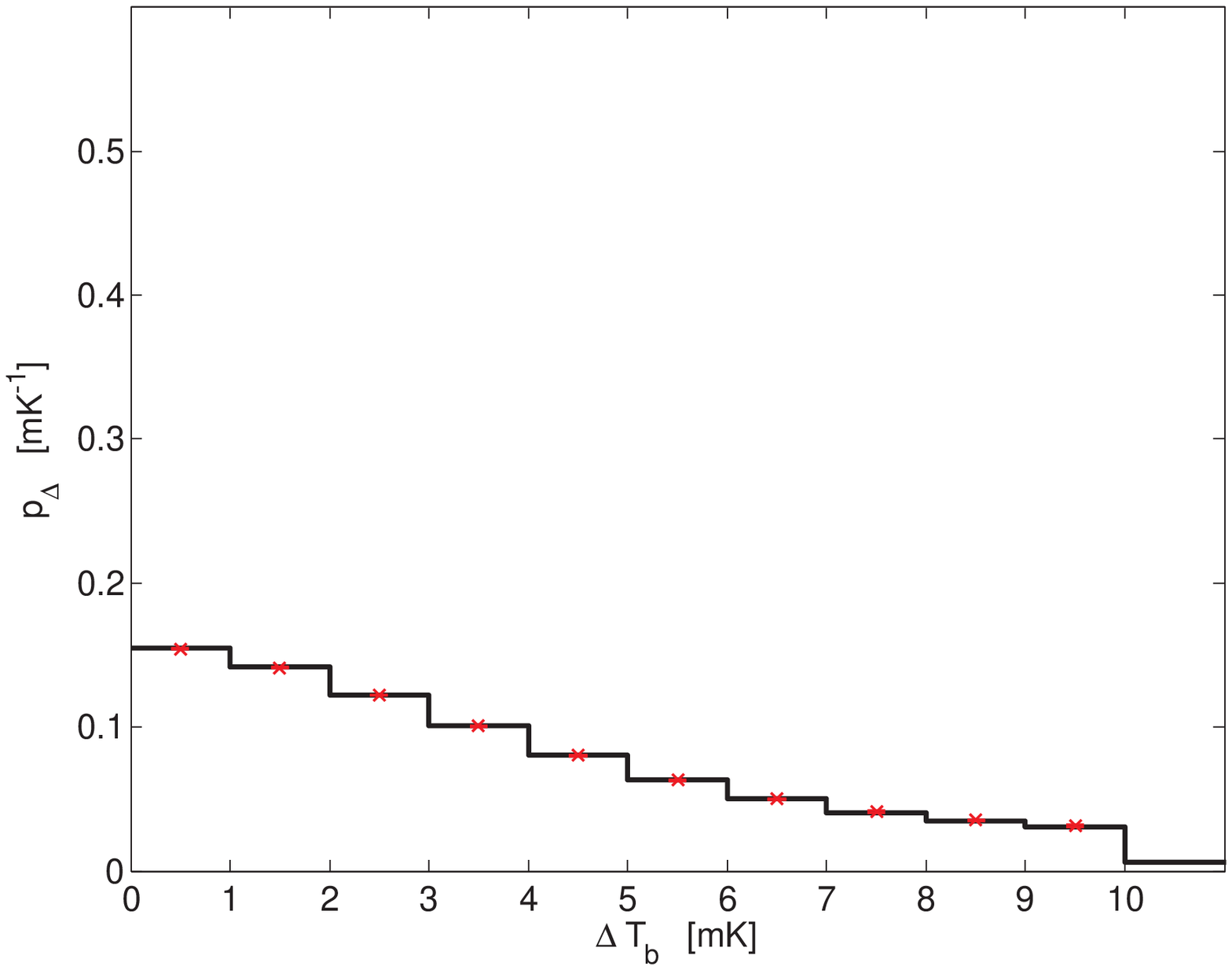}
\label{stair_figure-z82-rI6:SNf10}}
\caption{Same as Figure~\ref{stair_figure-z82-rI3:globfig} but for the
$r_{\rm mid}=119.5$ Mpc bin.}
\label{stair_figure-z82-rI6:globfig}
\end{figure*}

Near the midpoint of reionization, we illustrate the results for the
same three separation bins, in
Figures~\ref{stair_figure-z82-rI3:globfig},
\ref{stair_figure-z82-rI5:globfig}, and
\ref{stair_figure-z82-rI6:globfig}. The results here are similar,
in that useful measurements require MWA/10 errors for $r_{\rm
mid}=16.2$ Mpc, MWA/2 for $r_{\rm mid}=61.4$ Mpc, and just MWA at the
highest separation.  Here again the $r_{\rm mid}=$ 8.3 Mpc bin (not
shown) is similar to the 16.2 Mpc case, except for significantly less
accurate (though still useful) measurements for MWA/10. 

\begin{figure*}
\centering
\subfloat[Subfigure 1 list of figures text][MWA]{
\includegraphics[width=0.31\textwidth]
{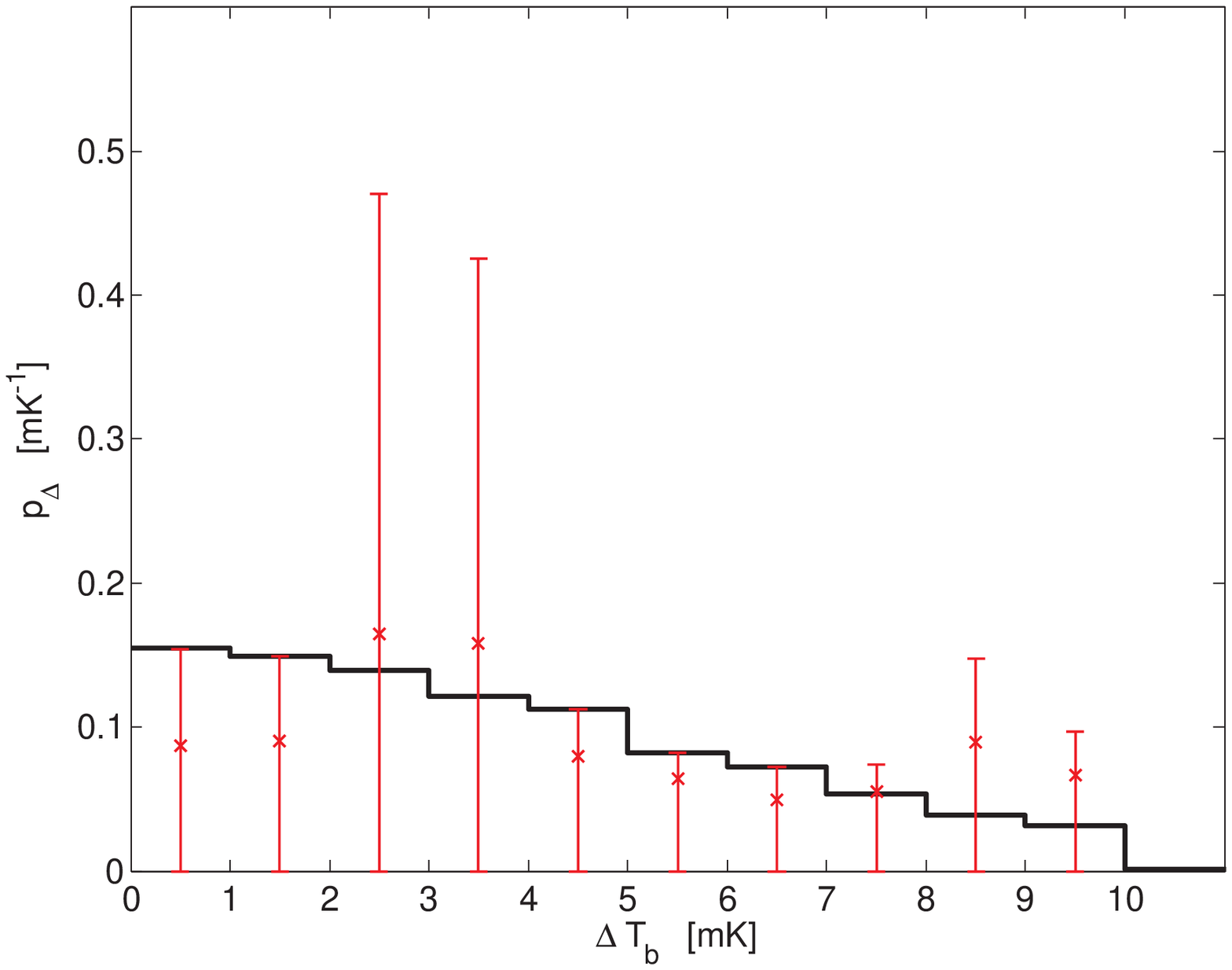}
\label{stair_figure-z101-rI5:SNf1}}
\quad
\subfloat[Subfigure 1 list of figures text][MWA/2]{
\includegraphics[width=0.31\textwidth]
{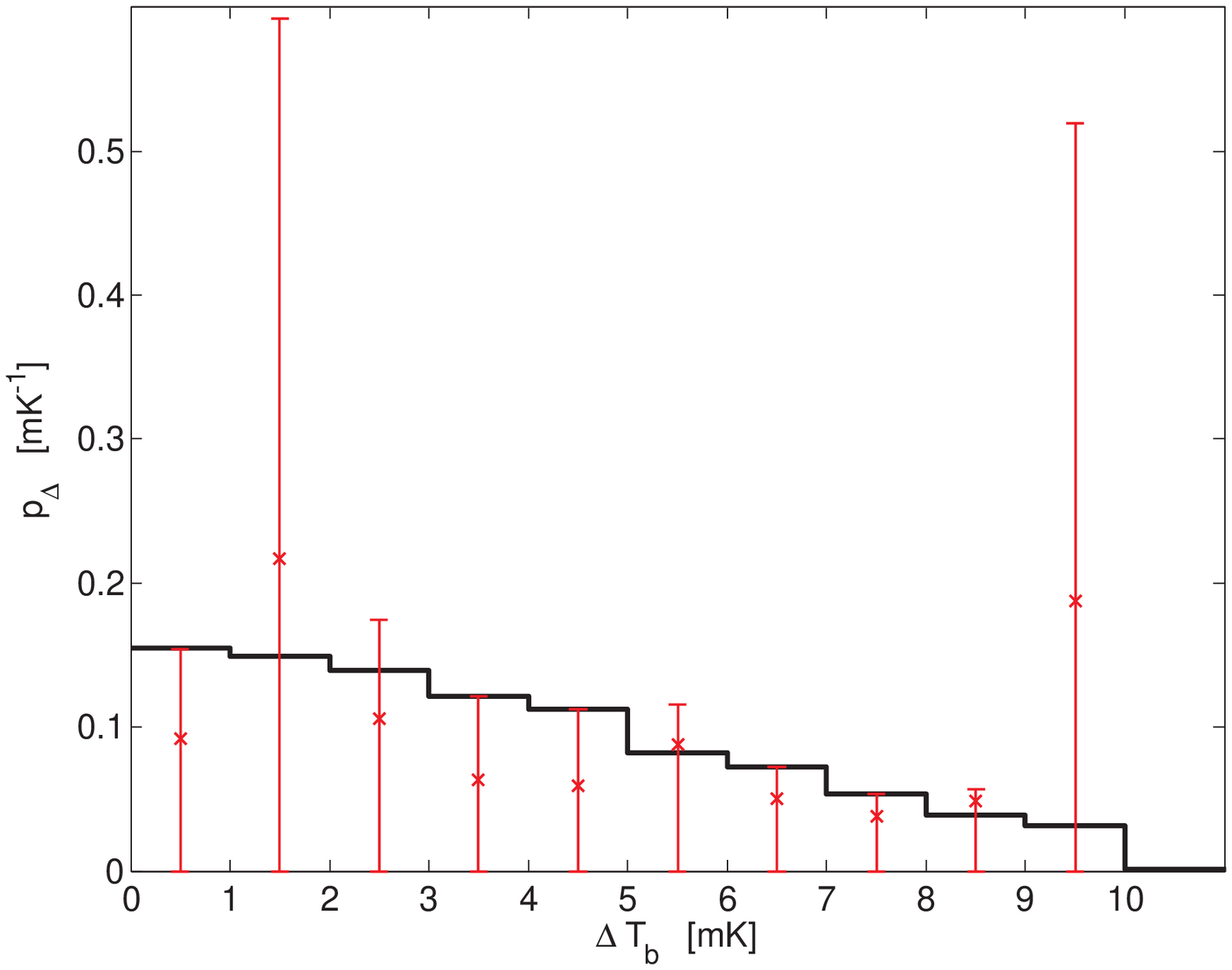}
\label{stair_figure-z101-rI5:SNf2}}
\quad
\subfloat[Subfigure 2 list of figures text][MWA/10]{
\includegraphics[width=0.31\textwidth]
{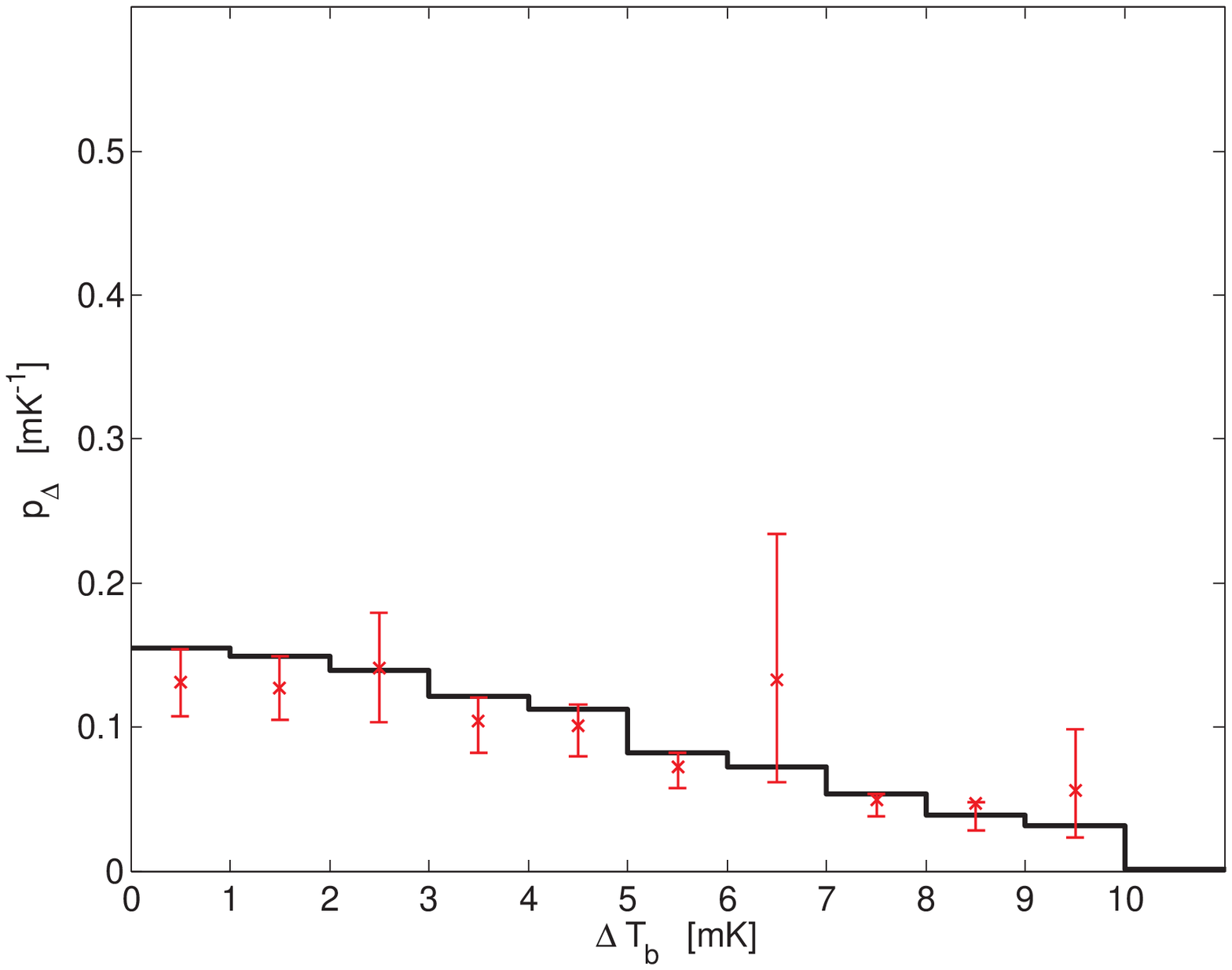}
\label{stair_figure-z101-rI5:SNf10}}
\caption{Same as Figure~\ref{stair_figure-z69-rI5:globfig}  
(i.e., $r_{\rm mid}=61.4$ Mpc) but at $z=10.1$.}
\label{stair_figure-z101-rI5:globfig}
\end{figure*}

\begin{figure*}
\centering
\subfloat[Subfigure 1 list of figures text][MWA]{
\includegraphics[width=0.31\textwidth]
{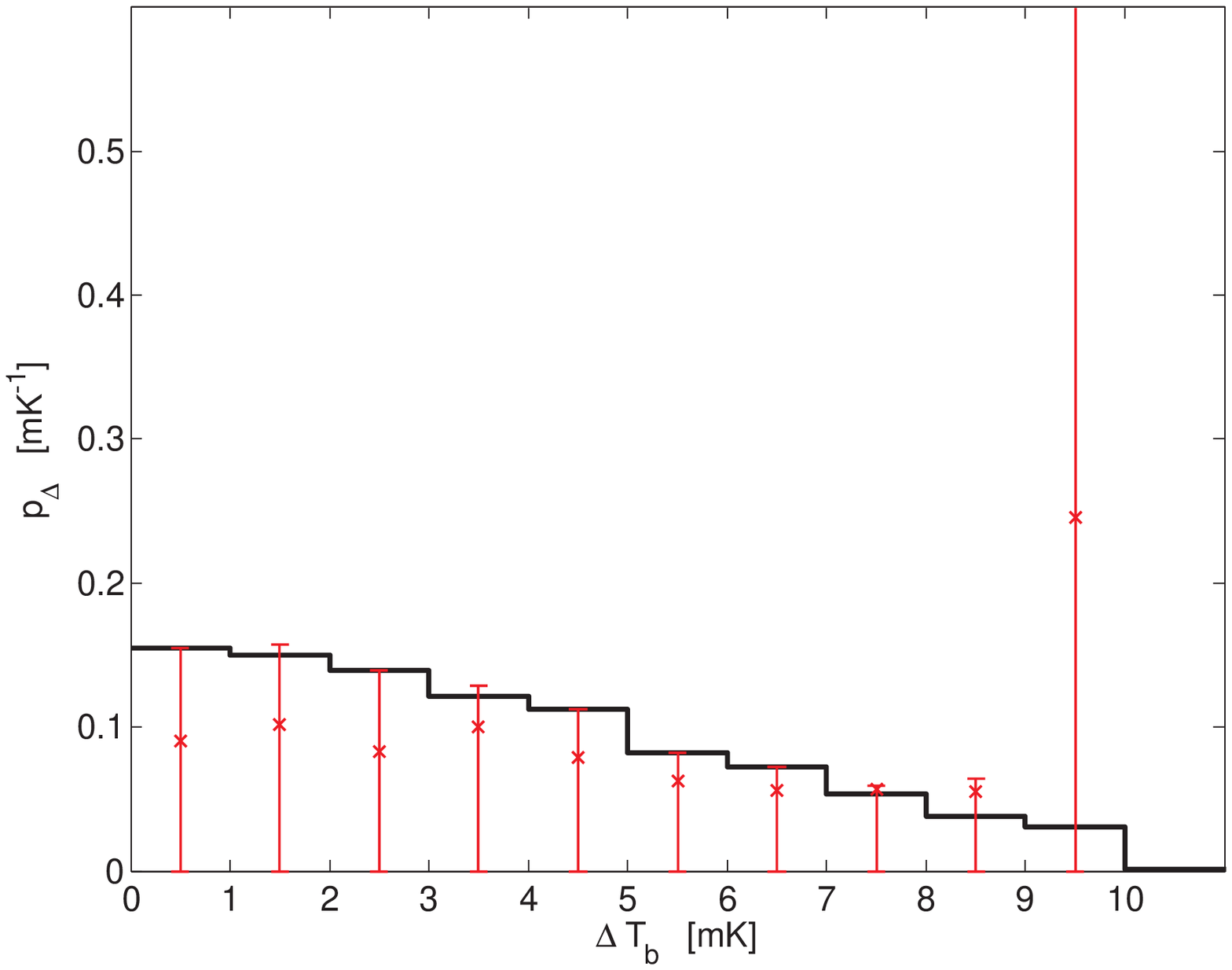}
\label{stair_figure-z101-rI6:SNf1}}
\quad
\subfloat[Subfigure 1 list of figures text][MWA/2]{
\includegraphics[width=0.31\textwidth]
{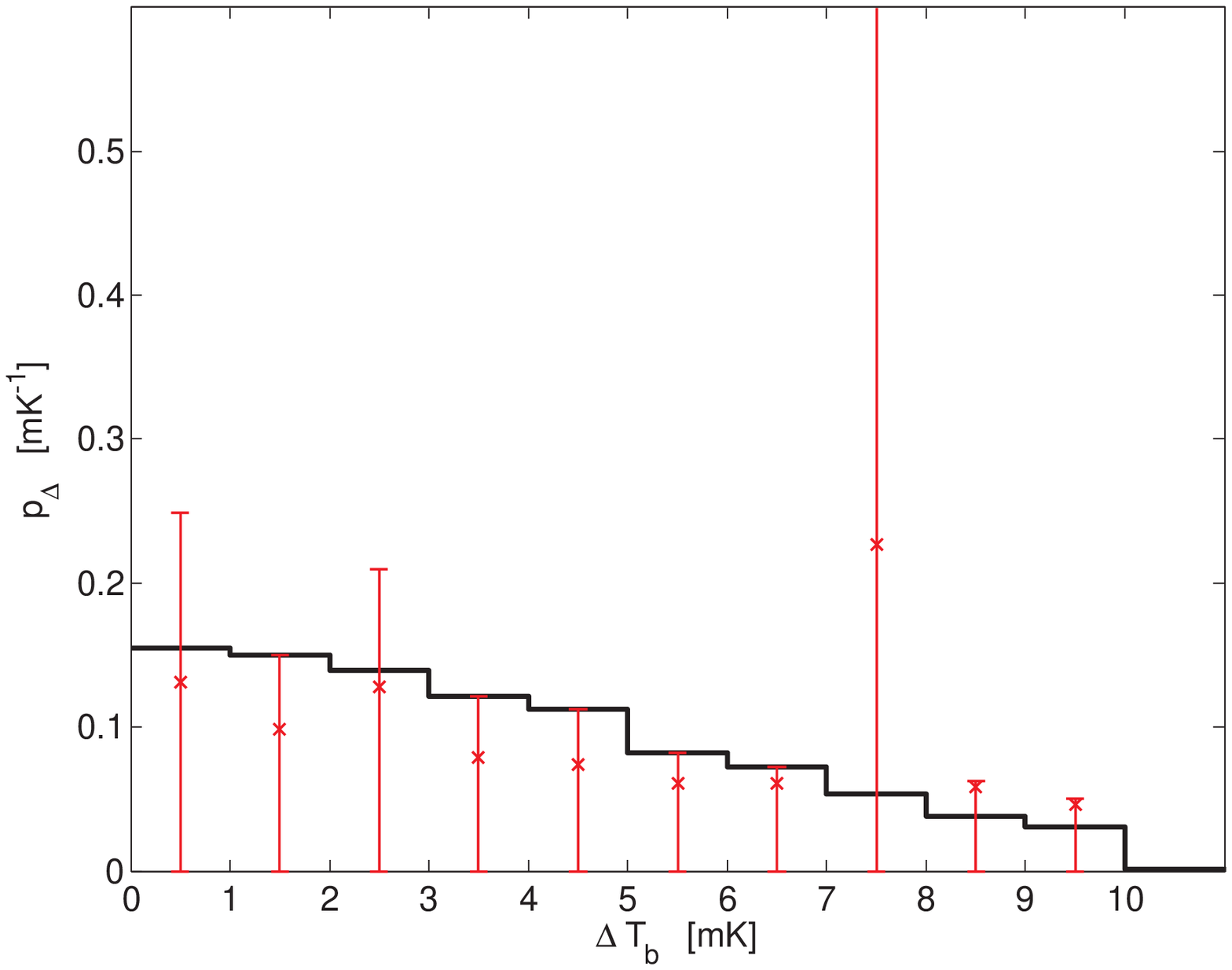}
\label{stair_figure-z101-rI6:SNf2}}
\quad
\subfloat[Subfigure 2 list of figures text][MWA/10]{
\includegraphics[width=0.31\textwidth]
{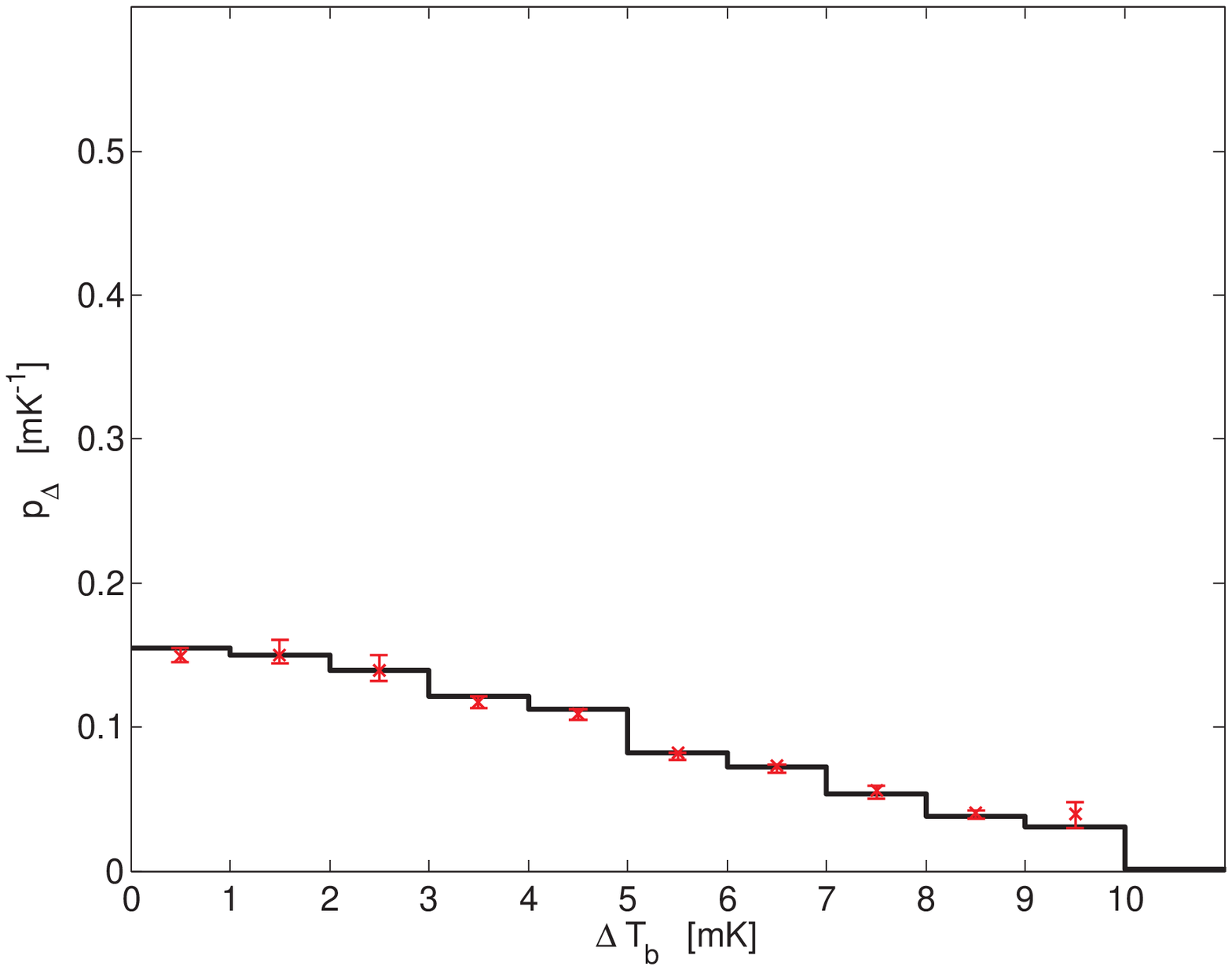}
\label{stair_figure-z101-rI6:SNf10}}
\caption{Same as Figure~\ref{stair_figure-z101-rI5:globfig} but for the
$r_{\rm mid}=119.5$ Mpc bin.}
\label{stair_figure-z101-rI6:globfig}
\end{figure*}

Finally, at a higher redshift when reionization is still in its early
stages, measurements are more difficult so we show only the two
highest distance bins, 61.4
(Figure~\ref{stair_figure-z101-rI5:globfig}) and 119.5 Mpc
(Figure~\ref{stair_figure-z101-rI6:globfig}). At this redshift, only
MWA/10 errors allow useful constraints on $p_\Delta$, in particular
giving a reasonable measurement at 119.5 Mpc. 

We conclude that with the 10-bin model, one-year MWA observations can
give a rough measurement of the difference PDF only at the highest
separation, and during mid-to-late reionization. Four-year
observations would substantially decrease the errors on the highest
separation bin, and give some constraints on the lower, $r_{\rm
mid}=61.4$ Mpc bin. However, only with MWA/10 thermal noise, i.e.,
with next generation radio array experiments, will it be possible to
recover the shape of the difference PDF across most distances, and
thus directly constrain the ionization correlation length (or bubble
size) with few assumptions needed. 

\begin{figure}
\centering
\includegraphics[width=1\columnwidth]
{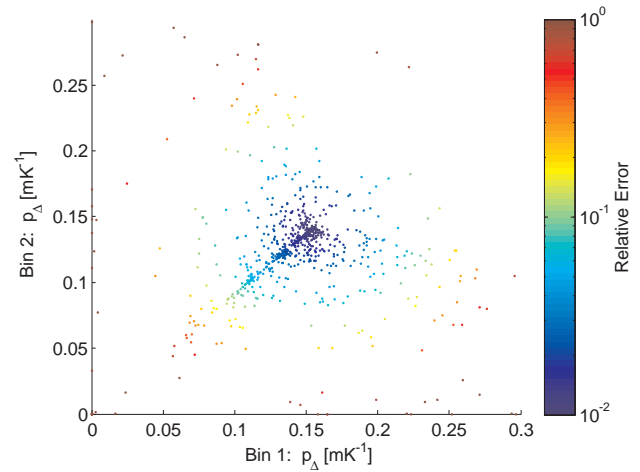}
\caption{Fits of the first 2 bins in the 10-bin model, 
at $z=8.2$, with $r_{\rm mid}=61.4$ comoving Mpc, for MWA/2 noise.
Bin 1 is $\Delta T =0$ to 1 mK, while Bin 2 is $\Delta T =1$ to 2 mK.
We color code the fit values for each of the 1,000 instances of mock
observational data by the root-mean-square of the relative errors in
Bin 1 and Bin 2. The mean fit and 1-$\sigma$ errors of all 10 bins can
be seen in Figure~\ref{stair_figure-z82-rI5:SNf2}. }
\label{PlotDegeneracy2D}
\end{figure}

With multiple parameters in the 10-bin model, the results are driven
by degeneracies between the probability fit values of various bins.
We plot an example of this in Figure~\ref{PlotDegeneracy2D}, for the
first two bins out of the ten. Since it is hard to distinguish
neighboring bins, a significant fraction of the values lie along a
diagonal line that illustrates a strong positive correlation. A few
poor fits lie on the axes, since the fit parameters for each $\Delta
T_b$ bin denote probabilities and were constrained to non-negative
values in the chi-squared minimization. Thirty-three fit values (out
of 1,000) lie outside the boundaries shown in the figure.

\section{Conclusions}

We have studied the expected errors in reconstructing the difference
PDF of the 21-cm brightness temperature during cosmic reionization.
We have shown how to perform a maximum likelihood fit to a model of a
binned difference PDF, and applied it to mock observational data
for a realistic field of view and a range of thermal noise levels. 

Previous work shows that the difference PDF during reionization should
display a strongly evolving shape that, if measured, can be used to
probe both the mean ionization history and the typical size of the
ionized bubbles. Early in reionization, the difference PDF still
resembles the Gaussian shape driven by density fluctuations, but it
later flattens and develops a two-peak structure, including a peak at
$\Delta T_b = 0$ due to jointly-ionized pixel pairs, and a peak at
$\Delta T_b \sim 10-15$ mK due to the temperature difference between
an ionized pixel and one that is still mostly neutral. The difference
PDF reaches an asymptotic form at large pair separations (where the
two pixels in the pair are essentially independent); the distance
below which it substantially changes shape is a measure of the
correlation length (of density, early on, and mainly of ionization
during the later stages of reionization).

We have found that a conservative approach that attempts only to
reconstruct the first bin of the difference PDF (which is related to
the ionization correlation function) should yield highly accurate
measurements. Even one-year MWA data should suffice for seeing the
signature of the end of reionization in the difference PDF, and for
probing large separations at earlier times.  Four-year data should
improve things markedly, typically decreasing errors by about a factor
of 4 (rather than the usual 2), since decreased noise helps remove
some of the partial degeneracies inherent in what is essentially an
attempt to deconvolve the noisy difference PDF. A second generation
experiment should be able to probe a wide range of separations during
most of the reionization era, assuming that reionization ends at $z
\sim 7$ and not much earlier. We note that measuring the 1-bin model
(i.e., a single parameter) is qualitatively similar in difficulty to
measuring the correlation function (or equivalently the power
spectrum), which is essentially equivalent to measuring a single
number (the variance) from the difference PDF in each bin of
separation distance.

Given these results with a 1-bin model, we have also considered a more
ambitious attempt to measure the detailed shape of the difference PDF
over ten bins. We found that with the 10-bin model, one-year MWA
observations can give a rough measurement of the difference PDF only
at the highest separation, and during mid-to-late
reionization. Four-year observations can give some improved
constraints, but only a second-generation radio array will make it
possible to recover the detailed shape of the difference PDF across a
range of distances and reionization stages.

We note that while we have only gone up to the bin of separation
distance centered at $r=120$ Mpc, it would be useful to measure bins
at even larger separations. Theoretically, $p_\Delta$ should be pretty
much constant with $r$ at such large distances, since the two voxels
of a pair are essentially independent of each other, but
observationally the number of pairs keeps rising with distance. The
total number of pairs in the MWA field of view, given the pixel size
of 2.9 Mpc, is $\sim 3 \times 10^{15}$ at $z=8$. Thus, the number of
pairs available for measuring the large-separation difference PDF is
potentially 100 times larger than the value we assumed, which was
already high compared to the available numbers at smaller
separations. While a measurement of the large-separation difference
PDF would not probe correlation functions, it would probe the cosmic
mean ionized fraction and, essentially, the one-point PDF (which
independently describes each point of the pair) at an exquisite
precision. 

Now that we have shown that measurements of the difference PDF are
quite promising relative to the expected thermal noise, the next
challenge is to consider similar statistics in the presence of
realistic foreground residuals and other systematic errors. In
particular, systematic errors that vary across the field of view might
make it in practice difficult to include the just-mentioned
wide-separation pairs.

\bigskip
\section*{Acknowledgments.}
This work was supported by Israel Science Foundation grant 823/09. TP
was also supported by the Hertz Foundation.

\bibliographystyle{mn2e}
\bibliography{references}

\end{document}